\documentclass[twoside]{article}
\usepackage{amsmath}

\usepackage{titlesec,graphicx,amssymb,amstext,amsmath,mathtools,amsthm}
\usepackage{natbib}
\usepackage[dvipsnames]{xcolor}

\usepackage{blindtext} 

\usepackage[most]{tcolorbox}

\usepackage[sc]{mathpazo} 
\usepackage[T1]{fontenc} 
\usepackage{microtype} 

\usepackage[english]{babel} 

\usepackage[hmarginratio=1:1,top=32mm,columnsep=20pt]{geometry} 
\usepackage[hang, small,labelfont=bf,up,textfont=it,up]{caption} 
\usepackage{booktabs} 

\usepackage{lettrine} 

\usepackage{enumitem} 
\setlist[itemize]{noitemsep} 

\usepackage{abstract} 

\usepackage{titlesec} 

\usepackage{fancyhdr} 
\usepackage{titling} 

\usepackage{hyperref} 

\usepackage{graphicx}
\usepackage{subcaption}

\theoremstyle{plain}
\newtheorem{theorem}{Theorem}[section]
\newtheorem{lemma}[theorem]{Lemma}

\newtheorem{open problem}[theorem]{Open problem}
\theoremstyle{remark}
\newtheorem{remark}[theorem]{Remark}
\theoremstyle{definition}

\newcommand{\xqedhere}[2]{%
  \rlap{\hbox to#1{\hfil\llap{\ensuremath{#2}}}}}
  
\providecommand{\keywords}[1]
{
  \small	
  \textbf{\textit{Keywords---}} #1
}

\pretitle{\begin{center}\Huge\bfseries} 
\posttitle{\end{center}} 
\title{Species Area Relationship (SAR): Pattern Description with Geometrical Approach\thanks{This work is done within 2017-2019. On the date of publishing on ArXiv, \textbf{Saeid Alirezazadeh} is with C4 - Cloud Computing Competence Centre (C4-UBI), Universidade da Beira Interior, Rua Marqu\^{e}s d'\'{A}vila e Bolama, 6201-001, Covilh\~{a}, Portugal, acknowledge the support given by operation Centro-01-0145-FEDER-000019 - C4 - Centro de Compet\^{e}ncias em Cloud Computing, cofinanced by the European Regional Development Fund (ERDF) through the Programa Operacional Regional do Centro (Centro 2020), in the scope of the Sistema de Apoio \`{a} Investiga\c{c}\~{a}o Cientif\'{i}ca e Tecnol\'{o}gica - Programas Integrados de IC\&DT. \textbf{Khadijeh Alibabaei} is with C-MAST Center for Mechanical and Aerospace Science and Technologies, University of Beira Interior, Deparment of Electromechanical Engineering 6201-001, Covilh\~{a}, Portugal acknowledge the support given by the project Centro-01-0145-FEDER000017-EMaDeS-Energy, Materials, and Sustainable Development, co-funded by the Portugal 2020 Program (PT 2020), within the Regional Operational Program of the Center (CENTRO 2020) and the EU through the European
Regional Development Fund (ERDF). Funda\c{c}\~{a}o para a Ci\^{e}ncia e a Tecnologia (FCT—MCTES) also provided financial support via project UIDB/00151/2020 (C-MAST).}} 
\author{%
\textsc{Saeid Alirezazadeh} \\[1ex] 
\normalsize CIBIO, InBIO, CEABN, Lisboa, Portugal \\ 
\normalsize \href{mailto:saeid.alirezazadeh@gmail.com}{saeid.alirezazadeh@gmail.com} 
\and 
\textsc{Khadijeh Alibabaei}\thanks{Corresponding Author}\\
\normalsize Dep. Matematica, FCUP, Porto, Portugal \\ 
\normalsize \href{mailto:f.alibabaee@gmail.com}{f.alibabaee@gmail.com} 
\and 
\textsc{Stephen P. Hubbell}\\
\normalsize Department of Ecology and Evolutionary Biology, \\
\normalsize University of California, Los Angeles, CA 90095}
\date{} 
\renewcommand{\maketitlehookd}{%
\begin{abstract}
Several formulations are describing the pattern of species-area relationship, log-log linear, semi-log linear, among others. These patterns mainly explain the species-area relationship for large areas, and for the small area, they provide significant differences from real data. We consider the geometric position of individuals of species, and base on that, we find the probability of observing at least one individual of the species. We apply a translation of the well-studied problem of mixed salt-water in a tank to describe the formula of SAR. For a rectangular sample area the species-area relationship follows the pattern, with some simplification, $S=c|A^{\beta}+a|^z$, where $S$ is the number of species in the area of size $A$ and $a,c,z$, and $\beta$ are constants with $z<1$ and $\beta\leq1$. We also show how the constant $z$ relates to some macroecological patterns, namely spatial aggregation, percentage of area coverage, and the core-satellite model. We exemplify our method using data on tropical tree species from a 50ha plot in Barro Colorado Island (BCI), Panama, using all individuals.
\end{abstract}
}

\begin{document}

\maketitle

\textbf{Data availability statement:}  The data are openly available in Smithsonian Libraries (Smithsonian research online) at \url{https://doi.org/10.5479/data.bci.20130603}.

\section{Introduction}

Patterns of biological diversity are scale-dependent \cite{levin1992, Whittaker2001}. One of the most commonly used tools to describe the scaling of biodiversity remains the species-area relationship (SAR) \cite{Lomolino2000, Palmer1994, He2002}.

The SAR is one of the oldest and best-documented patterns in ecology. It quantifies the relationship between area and the number of species present in that area, i.e., as a function of spatial scale. It is also a key tool to understand the pattern of species diversity. The scale dependence of biodiversity, as reflected in SAR, represents the combined effects of statistical sampling and ecological processes \cite{rosenzweig1995}. Three main factors influence the shape and slope of SAR, species rarity \cite{Preston1962}, habitat heterogeneity \cite{rosenzweig1995}, and spatial population dynamics \cite{Hanski1997, Taylor1978}. It remained open how to explain a biological mechanism for the shape of SAR. 

As a brief history of earlier work on SAR, see \cite{Connor1979}, where they discussed three issues concerning the basis, use, and interpretation of species-area curves, respectively the uniqueness, being optimized, and biological interpretation. The most common patterns of SAR are described by \cite{Preston1962} and \cite{Gleason1922} as log-log linear and semi-log linear, respectively. Fitting SAR based on data from field studies tends slightly in favor of the power law, and the exponent has been shown to depend on environmental variables, e.g., latitude \cite{Drakare2006}. Plotting the number of species by area in log-log scale suggests that the power-law can only be used as a suitable fit if the area is sufficiently large. Similarly, plotting the number of species by the log of the area suggests that the semi-log linear fit is good, but again only after a sufficiently large area. This suggests that both formulas depend on the choice of the initial small area to start the fit and that smaller areas do not follow the patterns they formulated. Our approach removes the dependence on the choice of initial small area and provides an estimated formula to describe the entire SAR.

\cite{Arrhenius1921} showed that SAR always has a negative second derivative when plotted arithmetically. \cite{Gleason1922} described the SAR as a straight line on semi-log axes ($S=c'+z' \log(A)$). \cite{Preston1962} described the SAR as a power function of area ($S=cA^z$), based on his work on the log-normal species abundance distribution. \cite{Conceicao2014} used a statistical approach and fit the SAR by assuming the existence of a random noise parameter, where the fitting is the sum of polynomials of the log of area, area, and reverse of area. Their formulation can be seen as a statistical generalized model. They considered species distributed in the area as compositions of normal, inverse Gaussian, and Gamma distributions. \cite{azaele:2015} proposed a scale-down method to obtain the SAR where for a given sample area and for a smaller scale they proposed the number of species on a smaller scale is a factor of the total number of species in the sample size. In their method, the factor at each scale can be obtained by solving some integrations of some Gamma distribution function. They provide a formula that may not have a simple solution without considering additional constraints on the parameters of the Gamma distribution functions. However, by what we will show in Section \ref{gama} even with the simplification, it has large complexity in computation at each scale. \cite{chrisholm:2016} described the shape of SAR for islands. \cite{Storch:2016} described the link between the SAR and other ecological patterns such as species abundance distribution, $\beta$-diversity, species richness, and productivity if we use geometric consideration of SAR. His work mainly shows the importance of geometrical consideration of the SAR. A more detailed history of SAR and mathematical expressions for SAR can be found in \cite{Tjorve:2003, Tjorve:2009, Tjorve:2012,whittaker:2020}. Note that different sampling circumstances can give rise to different SARs, \cite{Scheiner:2003}. Here, we are solely interested in: how does the number of species change when we sample nested areas of increasing size?

Note that, for a given sample data, all the preceding formulas of SAR can be a good fit only if the sample area and the initial sub-area are large enough, and for small ones, they have significant differences from real data. In most of the existing sample data, the semi-log linear fitting produces a negative number of species for small sub-areas, the log-log linear fitting produces a higher number of species than the ones existed in real for small sub-areas, and it has the problem of over-fitting due to insufficiency of data for a small area. By considering SAR as a continuous function, the result of \cite{Conceicao2014} can be interpreted as Weierstrass Approximation Theorem\footnote{firstly provided in the German language in \cite{Weierstrass1885} the English version can be found in any mathematical analysis books, for example, see \cite{Zeugmann2016} for the statement and a constructive proof}, states that every continuous function uniformly approximated as closely as desired by a polynomial function, but then the degree of that polynomial is what causes the overfitting. 

Since in the most existing sample data, the log-log linear fittings produce relatively better results than the semi-log linear, in our examples, we only compare with the log-log linear fitting. See Section V of Appendix \ref{AppA}.

Our goal is to describe a pattern which can be used for small sample area as well as for large area. Note that, throughout this manuscript, we explain SAR, which is built on a geometrical approach without considering any prior information. 

We exemplify our method using data on tropical tree species from a 50ha plot in Barro Colorado Island (BCI), Panama, using all individuals \cite{Hubbell2005, Hubbell1999, condit1998, condit2019}. We will use species codes instead of species names in our examples; see \ref{tab1}.

\begin{table}
\caption{For BCI data, the correspondence species names for the species code we will use in our examples.}\label{tab1}
\centering
\begin{tabular}{ll}
Species Code & Species name\\
\midrule
aegipa &  Aegiphila panamensis \\
entesc &  Enterolobium schomburgkii \\
rinosy & Rinorea sylvatica \\
maytsc &	Maytenus schippii\\
anaxpa & Anaxagorea panamensis \\
cha2sc & Chamguava schippii \\
appuse & Appunia seibertii \\
bactc1 & Bactris coloniata \\
\bottomrule
\end{tabular}

\end{table}


\section{Preliminaries}

We say that a community is split when all species (except a very small number of species) are rare or very abundant. In other words, the species abundance distribution has a U-shape. Such communities are not in the interest of this manuscript, because in reality, we can only detect such a community if the sample size is very small.

\begin{figure}[!h]
\centering
\includegraphics[width=0.6\linewidth]{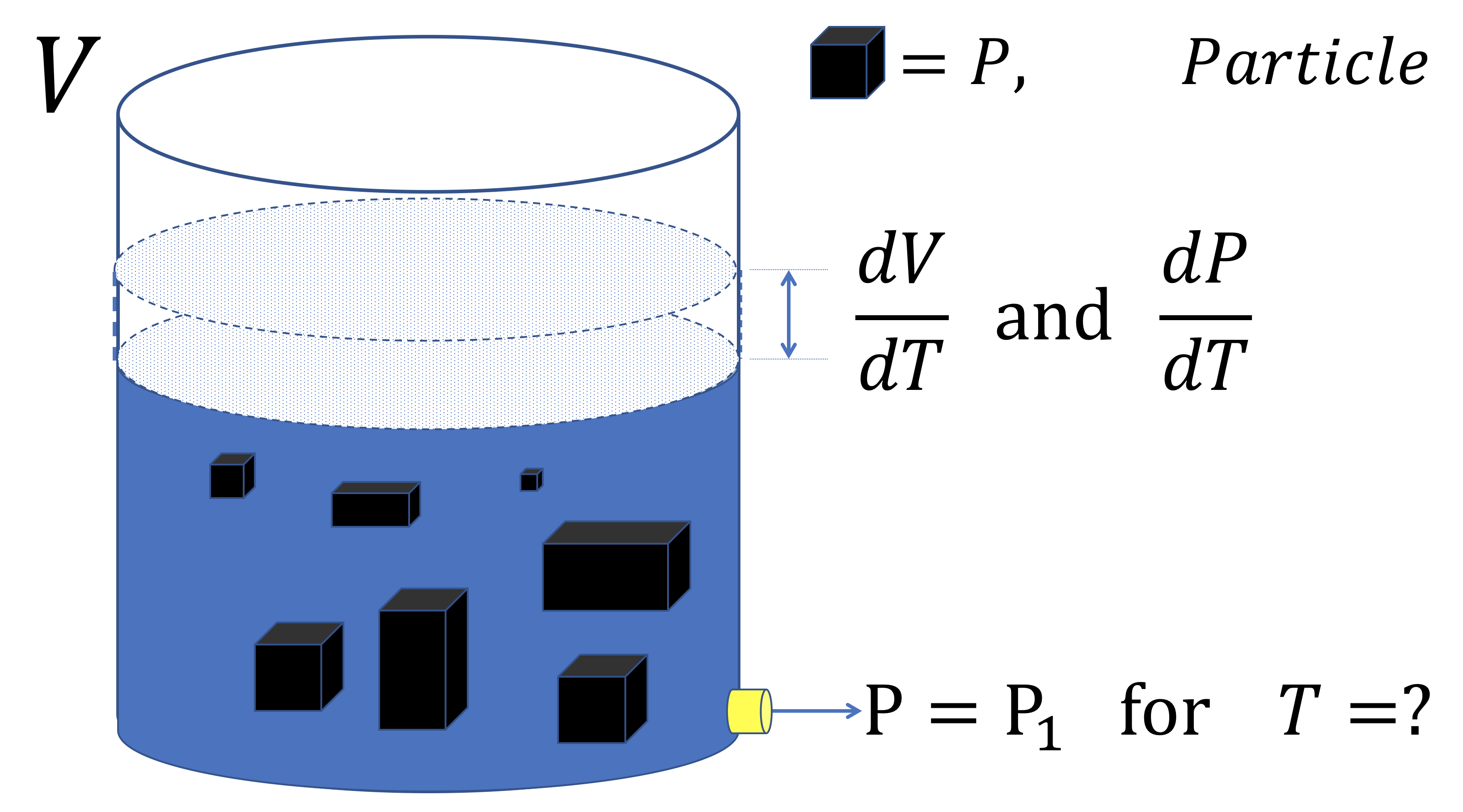}
\caption{Mixed salt-water in a tank problem.}
\label{fig1}
\end{figure}

Let $A=(X,Y)$ be a rectangular sample area, where $X$ and $Y$ are the width and height of the area, respectively, and it consists of the position of all individuals of each species. We must assume that all individuals are almost uniformly distributed in the sample area, or at least for a reasonably small sub-area $B$ size, each sample from the total area has a sub-area with size greater than or equal to the size of $B$ with no significantly different number of individuals. This allows us to consider the number of individuals to be linearly increasing as a function of subarea size, which is the case for our test data, BCI. To identify the functionality of SAR, i.e., to provide a function to describe the pattern of SAR, we proceed by finding the probability of observing at least one individual of a fixed species across scales, which can be viewed as the change in the number of individuals of a fixed species over the spatial scale. We then apply a translation of the mixed salt-water in a tank problem as follows: consider a tank with a fixed volume containing water and salt. The initial volume of water and salt is known and fixed. Assume that the tank has a hole and loses a fixed volume of water per unit of time. One of the problems that can be answered by the problem of mixed salt-water in a tank is: at what time will a certain amount of salt be lost from the tank (see Figure \ref{fig1}). 

As a simple example, suppose that the tank consists of $90$ Lt of water and $10$ kg of salt and that it loses $1$ Lt of the mixture per hour. The question is when the tank will lose $1$ kg of salt. Note that it is assumed that the salt is well mixed in the water. In general, instead of salt, the tank can be assumed to contain particles of different volumes and hence different weights. We attempt to translate this problem into SAR to answer this question, with the solution providing a descriptive answer to the formulation of SAR. We will explain later why we consider this case. We should note that we do not consider the entry of water or particles into the tank.

\subsection{How this translates to SAR}
Given a rectangular sample area $A=(X, Y)$, where $X$ and $Y$ are the width and height of the area, respectively, and it consists of the positions of all individuals of all species. We call a rectangular area $B=(X',Y')$ (which may lie within $A=(X,Y)$) a subarea of $A$, if the equality $\frac{X'}{X}=\frac{Y'}{Y}$ holds. To describe SAR, fix a subarea of the total area $A$ and then determine the number of species in that subarea. Since the position of the subarea is not fixed, a change in position can change the number of species and also the type of species. To remove the dependence on the position of the subarea, one can randomly select several positions for the subarea and then average the number of species with respect to the different positions. By this random placement, and as we explain later, the number of species associated with the subarea can be obtained by $p_1+\cdots+p_S$, where $S$ is the total number of species in $A$, and $p_s$ is the probability of observing at least one individual of species $s$ in the subarea. 

\begin{figure*}[!h]
\centering
\includegraphics[width=0.8\linewidth]{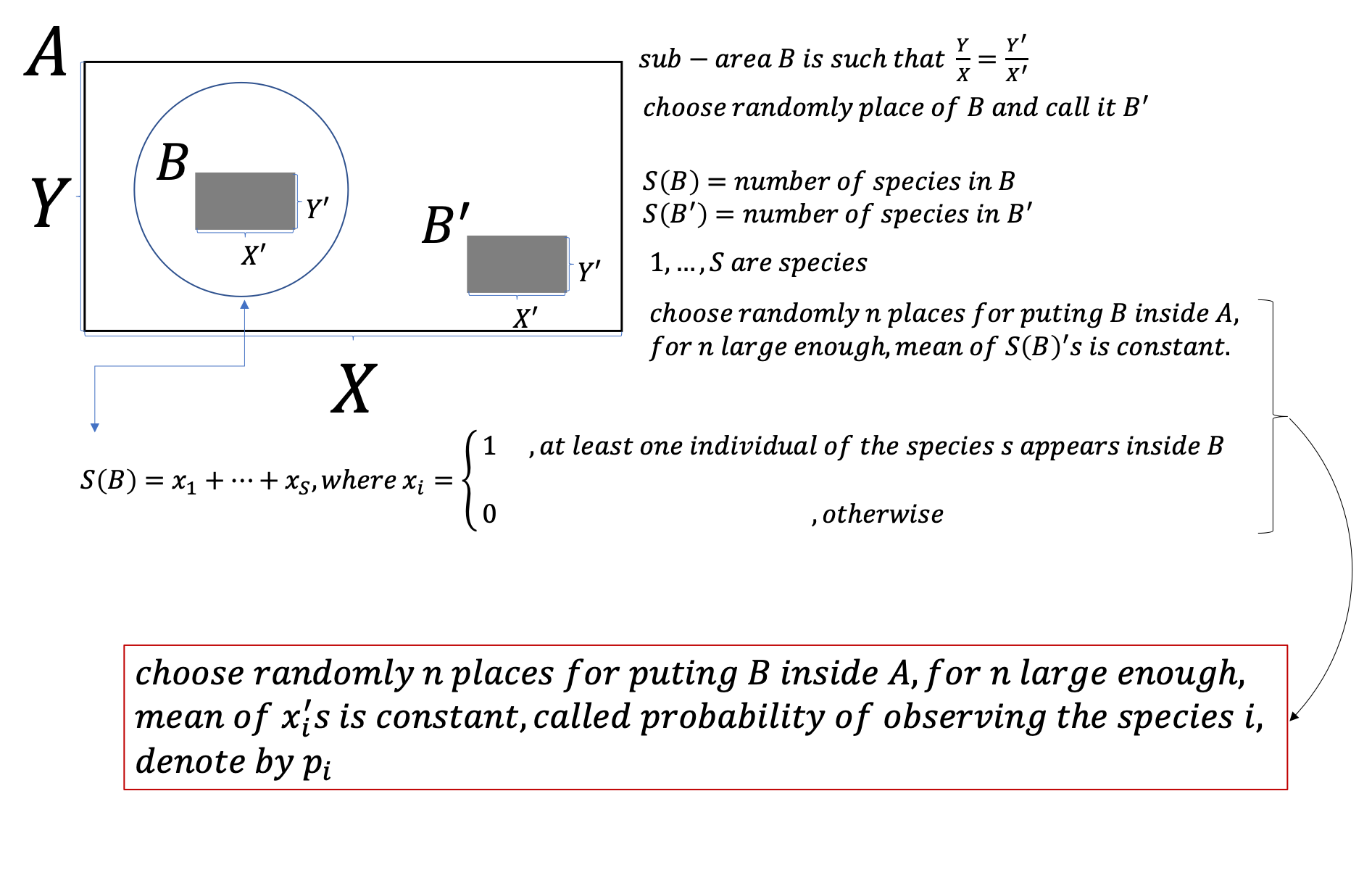}
\caption{An intuitive explanation for the number of species in a subarea is the sum of probabilities of observing species in the sub-area. $B$ and $B'$ are sub-areas of $A$. Both are the same proportion of the total area of $A$ but in different positions. Note that the index $S$ denotes the total number of species, and the index $i$ denotes the species $i$.}
\label{fig2}
\end{figure*}

Recall that for a given subarea of $A$, and by fixing its position, we can consider the species occurring in the subarea as a list $(x_1,\ldots,x_S)$, where $x_i$ for $i=1,\ldots, S$ is either one if at least one individual of species $i$ occurred in this subarea and zero otherwise. Thus, the total number of species in this subarea is equal to $x_1+\cdots+x_S$. Changing the position of the subarea may change this list. However, as we explain below, its mean converges by choosing subareas in infinitely many different positions; see Figure \ref{fig2}.

Fix one species and consider its individuals as salt in the tank. Consider the individuals of the other species as water. We know that the number of individuals as a function of the area follows the linear relation $N(A)=\rho A$, where $N(A)$ is the number of individuals in area $A$ and $\rho$ is a positive constant. Now suppose that the size of the subareas is time, i.e., if the sequences of subareas are $A_1,\ldots, A_n=A$, then $A_{(i+1)}-A_i$ is fixed for all $i$'s. That is, the amount of water lost from the tank per unit time is fixed. Since the individuals of a species are not uniformly distributed in the area (it is equivalent to considering that the salt is not dissolved in water), we consider the general formulation of the mixed salt-water problem. Moreover, we do not introduce any new individuals in the process, which is equivalent to not adding water or salt to the tank. We first proceed to formulate it for one species at a time, and then the pattern of SAR can be described as a solution to the sum of formulas for all species. See Appendix \ref{AppA}, as a way to obtain SAR for a given sample data in detail.

\begin{figure}[h!]
\centering
\includegraphics[width=0.4\linewidth]{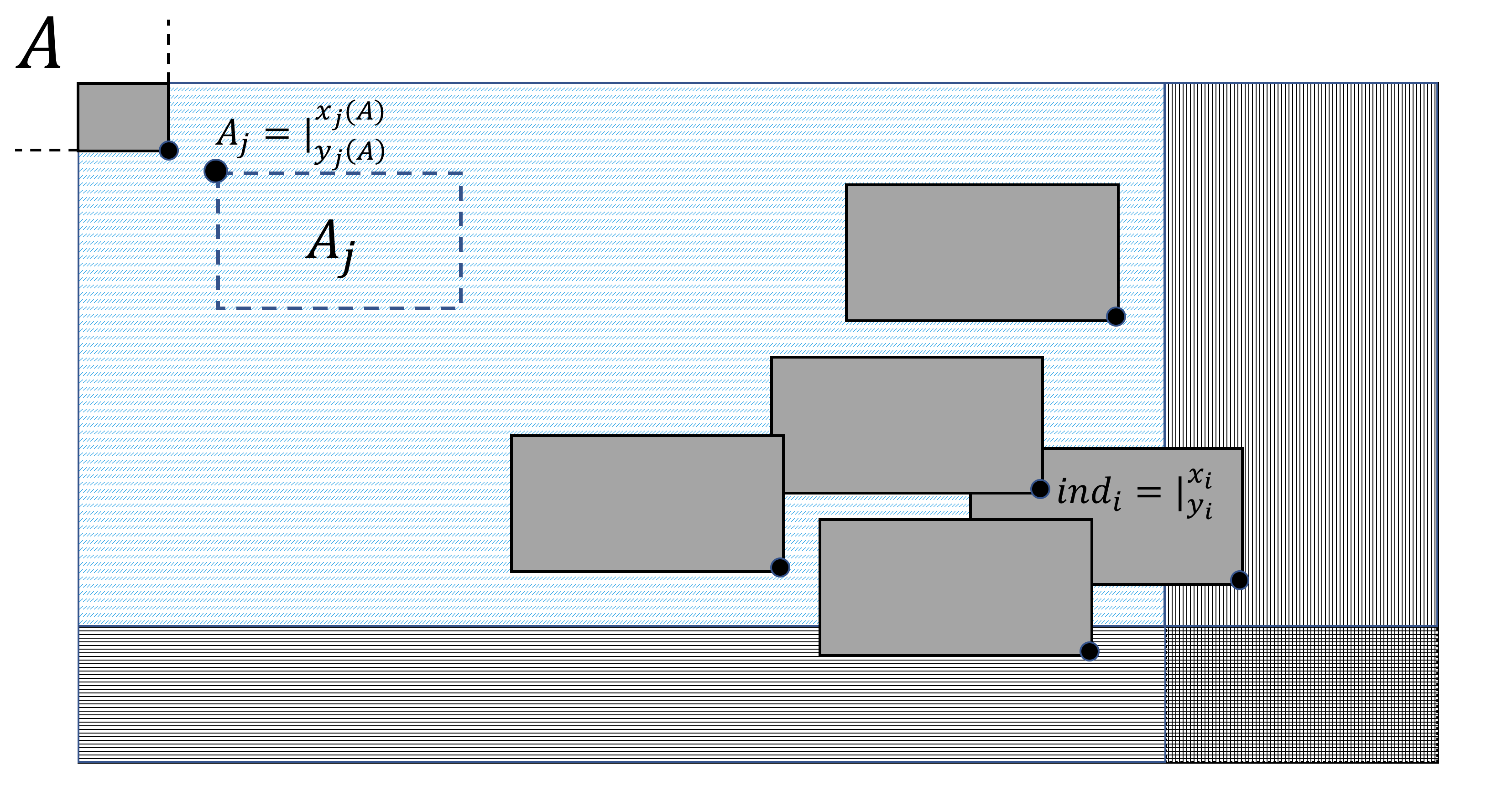}
\includegraphics[width=0.6\linewidth]{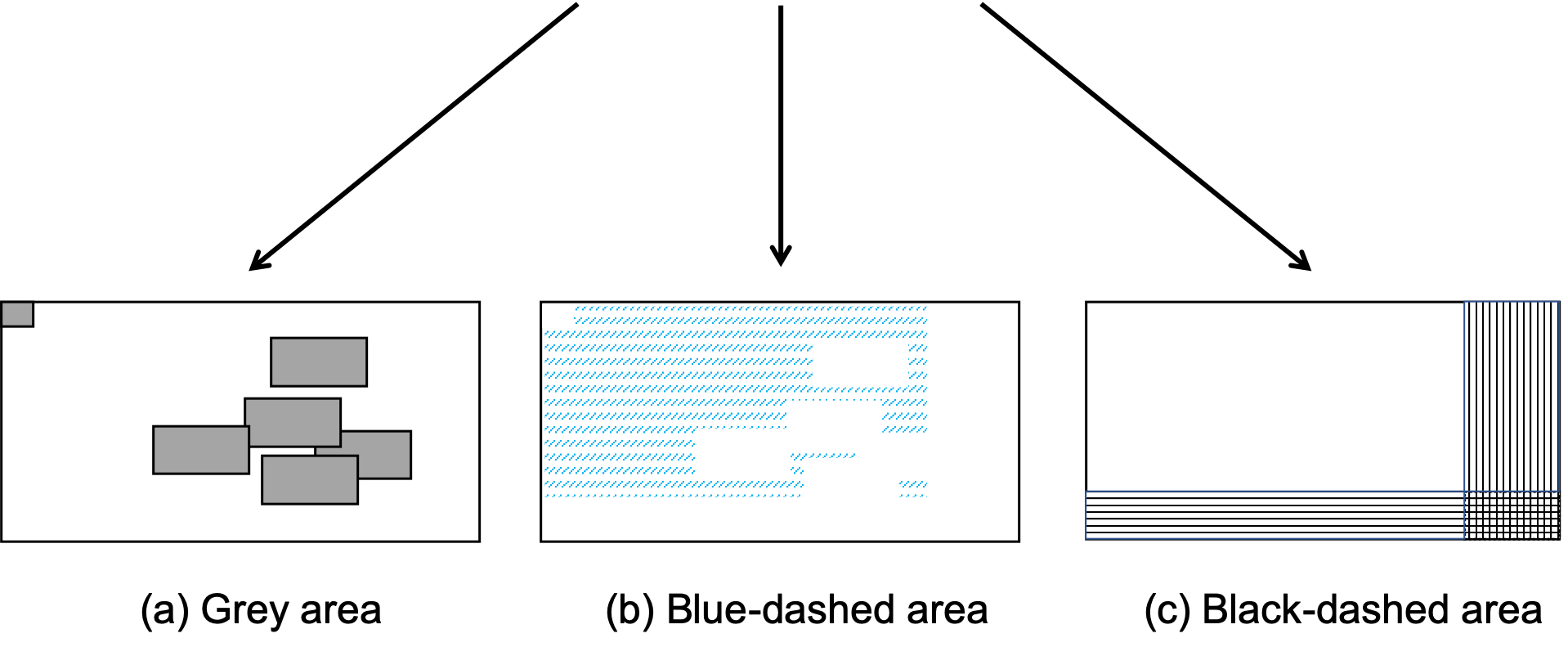}
\caption{For a given rectangular area and a fixed species, we show how the gray rectangles are the subareas associated with individuals of the species and the subareas associated with a given point in the area. $ind_i=(x_i,y_i)$ represents the position of the $i$-th individual of a fixed species $s$. For clarity, we decompose the large figure into three smaller subfigures. Subfigure (c) shows the black-dashed area; the black-dashed area is the strip on the right and bottom of the total area, it represents the positions in the total area $A$, so if we choose any point on it as the position of the upper left corner of the subarea $A_j$ we cannot fit the subarea $A_j$ into the area $A$. Subfigure (a) shows the gray area; the gray area represents the positions in the total area $A$, whereby choosing any point on them as the position of the upper left corner of the subarea $A_j$, at least one individual of the species $s$ can be observed. The rest of the positions in the total area of $A$ that are neither in the black-dashed area nor in the gray area is the blue-dashed area (subfigure (b)). By choosing any point on the blue-dashed area as the position of the upper left corner of subarea $A_j$, we can fit subarea $A_j$ into area $A$, but we cannot observe an individual of species $s$.}
\label{fig3}
\end{figure}

Any position of a subarea of size $A_j$ can be identified by the position of a point, which is the position of the point of the upper left corner of that subarea. For example, the rectangle with dashed borders in Figure \ref{fig3} identifies a particular position of the sub-area $A_j$ in the overall area $A$ by the single point on its upper left corner. This is because the width and height of a subarea are already known, and we only need to know the position of a single point in the subarea, from which the positions of the other points in the subarea can then be easily derived. If we fix a species $s$, we can assign a subarea to each of its individuals: If we fix an individual of species $s$, its associated subarea is the largest subarea of size at most $A_j$ that lies entirely within $A$, and this individual is located in its bottom-right corner (gray rectangles in Figure \ref{fig3}, see Appendix \ref{AppA} for details). In this way, we can directly determine the probability of observing a species in a subarea of a given size, rather than randomly selecting subareas infinitely many times and then taking their mean.

For a fixed species $s$ and a subarea $A_j$, the total area $A$ can be divided into three areas. 
\begin{itemize}
\item The black dashed area: it represents the positions in the total area $A$ such that if you choose any point on it as the position of the upper left corner of the subarea $A_j$, you cannot fit the subarea $A_j$ into the area $A$;
\item The gray area: it represents the positions in the total area $A$ whereby choosing any point on them as the position of the upper left corner of the subarea $A_j$, at least one individual of the species $s$ can be observed;
\item The blue-dashed area: it represents the positions in the total area $A$ such that by choosing any point on the blue-dashed area as the position of the upper left corner of the sub-area $A_j$, the sub-area $A_j$ can be fitted into the area $A$, but no individual of the species $s$ can be observed.
\end{itemize}

We have shown that for a fixed subarea $A_j$, the number of species in $A_j$, $S(A_j)$, is equal to the sum of probability of observing each of the species (see Appendix \ref{AppA}). Denote by $S(A_j)\mid_s$, the restriction of the number of species in $A_j$ to species $s$, i.e., ignore all other species except $s$. By this notation, $S(A_j)\mid_s$ is equal to the probability of observing at least one individual of species $s$ in subarea $A_j$.

It follows, 
\begin{equation*}
S(A_j )\mid_s=1-\frac{\text{Area of Blue dashed}}{A-\text{Area of Black dashed}}
\end{equation*}
which is equal to 
\begin{equation*}
\frac{\text{gray Area}\cap(A-\text{Area of Black dashed})}{A-\text{Area of Black dashed}}, 
\end{equation*}
which is the probability of observing at least one individual of species $s$ in the subarea of size $A_j$.

Now, denoting the subarea by $B$, the preceding can be written as $S(B)\mid_s=\frac{\varepsilon_s (B)}{\alpha(B)}$, where $\varepsilon_s(B)$ and $\alpha(B)$ are functions of $B$. If we denote the height and width of $A$ and $B$ respectively by $y_A$, $y_B$ and $x_A$, $x_B$, then $\alpha(B)=(y_A-y_B)(x_A-x_B)$, which is a strictly decreasing function of $B$ and as a function of subarea $\alpha(B)=A+B-2\sqrt{ AB }=(\sqrt{A}-\sqrt{B})^2$. The function $\varepsilon_s (B)$ refers to how individuals of the species disperse in the total area. Note that for each species, the corresponding function $\varepsilon_s (B)$ increases for small $B$ and decreases above a certain subarea, in other words, $\varepsilon_s (B)$ is a bell-shaped function, so we can denote it by a non-symmetric Gaussian function:
\begin{equation*}
\varepsilon_s (B)\approx f_s (B) \frac{1}{\sqrt{2\pi c_s}}\exp(\frac{-(B^{\beta_s}-b_s)^2}{2c_s}),
\end{equation*}
where $b_s$ and $c_s$ are constants and $f_s (B)\geq0$, is a positive function over $B> 0$. Because of the skewness of $\varepsilon_s (B)$, we can consider it as a constant multiple of a skew-(generalized) normal distribution function. The general formulation of such a bell-shaped function with a slight modification is:
\begin{align*}
\varepsilon_s (B)\approx&\alpha_s  \frac{1}{\sqrt{2\pi c_s}} \exp(\frac{-(B^{\beta_s}-b_s)^2}{2c_s})\\
&~~\left(1+\mathrm{erf}(\frac{e_s(B^{\beta_s}-b_s)}{\sqrt{2c_s}})\right),
\end{align*}
which means 
\begin{equation*}
f_s(B)=\alpha_s\left(1+\mathrm{erf}(\frac{e_s(B^{\beta_s}-b_s)}{\sqrt{2c_s}})\right),
\end{equation*}
see \cite{azzalini2014, DASILVAFERREIRA2011}. The function $\varepsilon_s (B)$ describes several properties of the species. If we consider $B\rightarrow0$, then the limit $\lim_{B\rightarrow0}\varepsilon_s(B)$ gives the frequency of the species $s$. Thus, $\varepsilon_s (0)$ refers to the frequency of species $s$.

The parameter $\beta_s$ induces a weight for the subarea of solid grey. It is a factor that enforces the regularization of the dispersal of individuals of a species in the total area. It acts as a compensation for the speed at which the gray area reaches its maximum value (the faster the gray area reaches its maximum value, the smaller $\beta_s$). By forcing $\beta_s=1$, we remove the weight of the gray subarea, and we can compare some properties of species with the same number of individuals. 

If we let $\beta_s=1$ hold for all species, in order to find the value of $e_s$, we need to know how fast the following area reaches its maximum
\begin{equation*}
\text{Area of solid grey}\cap(A-\text{Area of Black dashed}).
\end{equation*}
Whenever individuals of the species $s$ are closer to the center (core-satellite), the value of $e_s$ is larger. Also, the parameter $c_s$ reflects the property of the individuals of the species. If the individuals of species $s$ are distributed over the whole area, then the value of $c_s$ is small, and if either the individuals of species $s$ are clustered or the species $s$ is rare, then it is large. See Appendix \ref{AppC} for how individuals of a species can be identified as clusters at a given size.

\subsection{Applied to real data}

\begin{figure}[!h]
\centering
\includegraphics[width=\linewidth]{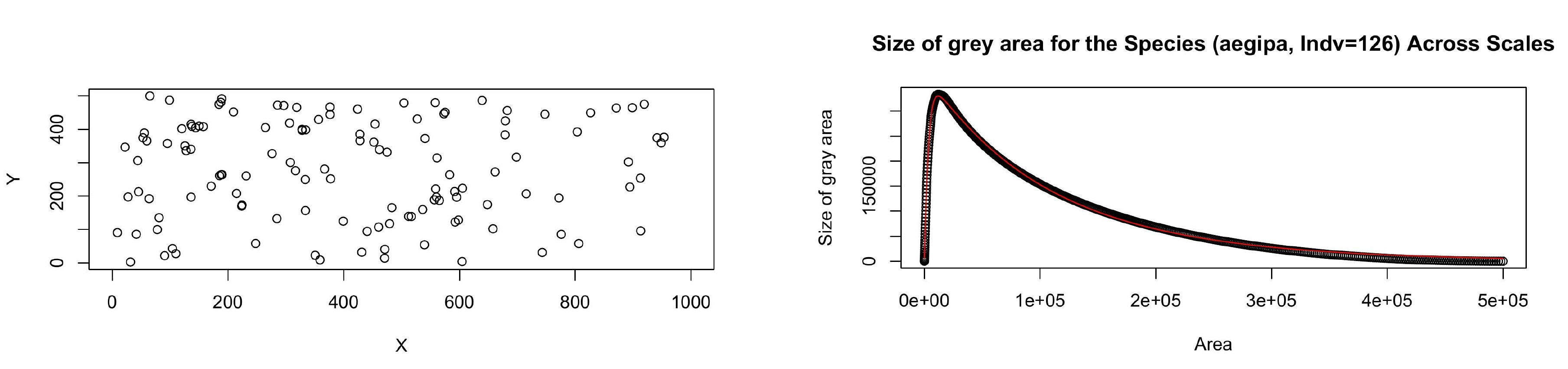}
\includegraphics[width=\linewidth]{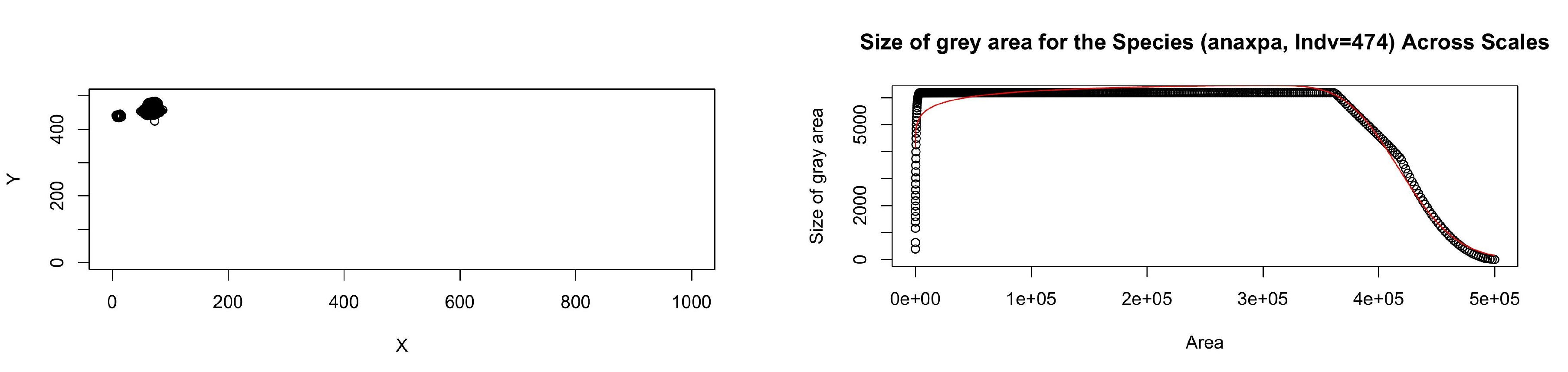}
\caption{For the BCI data, the left figures show how the individuals of the species aegipa and anazpa are distributed over the total area of 50 ha, and the right figures show the size of their gray area as a function of the size of the subareas. The red lines are their fits with the parameters in Table \ref{tab2}.}
\label{fig4}
\end{figure}

\begin{figure}[tbp]
\centering
\includegraphics[width=\linewidth]{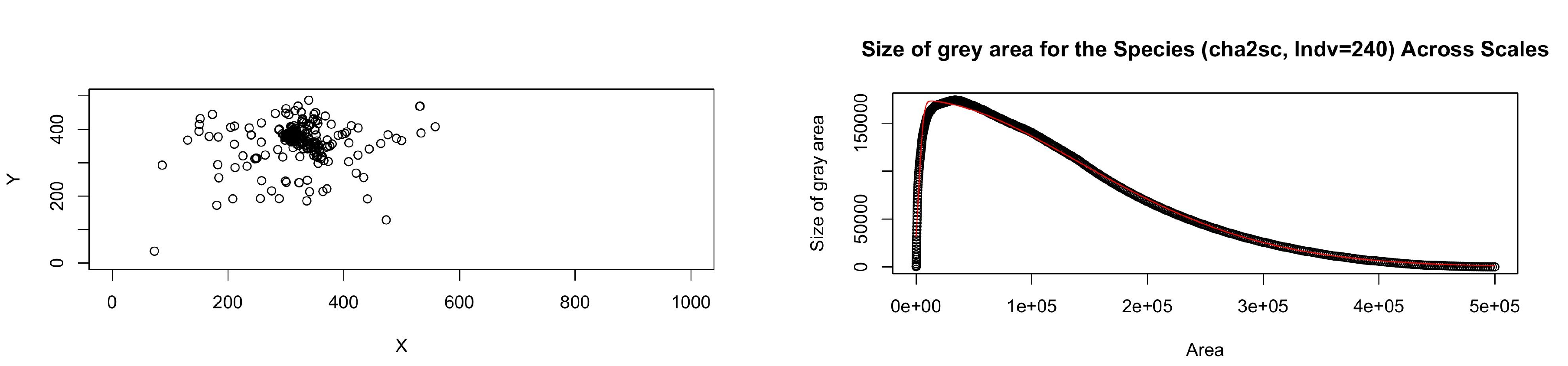}
\includegraphics[width=\linewidth]{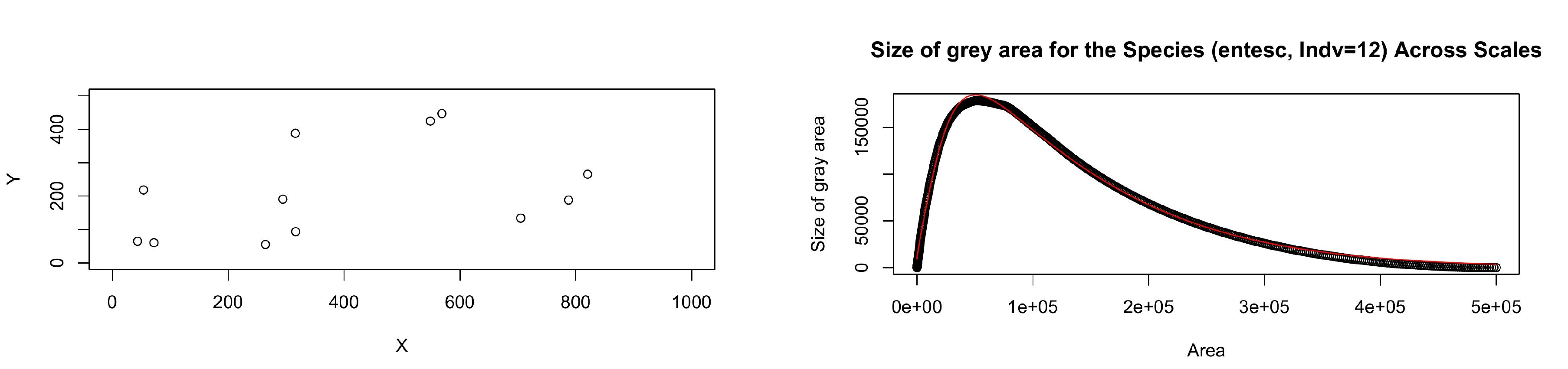}
\caption{For the BCI data, the left figures show how the individuals of the species cha2sc and entesc are distributed over the total area of 50 ha, and the right figures show the size of their gray area as a function of the size of the subareas. The red lines are their fits with the parameters in Table \ref{tab2}.}
\label{fig5}
\end{figure}

\begin{figure}[!h]
\centering
\includegraphics[width=\linewidth]{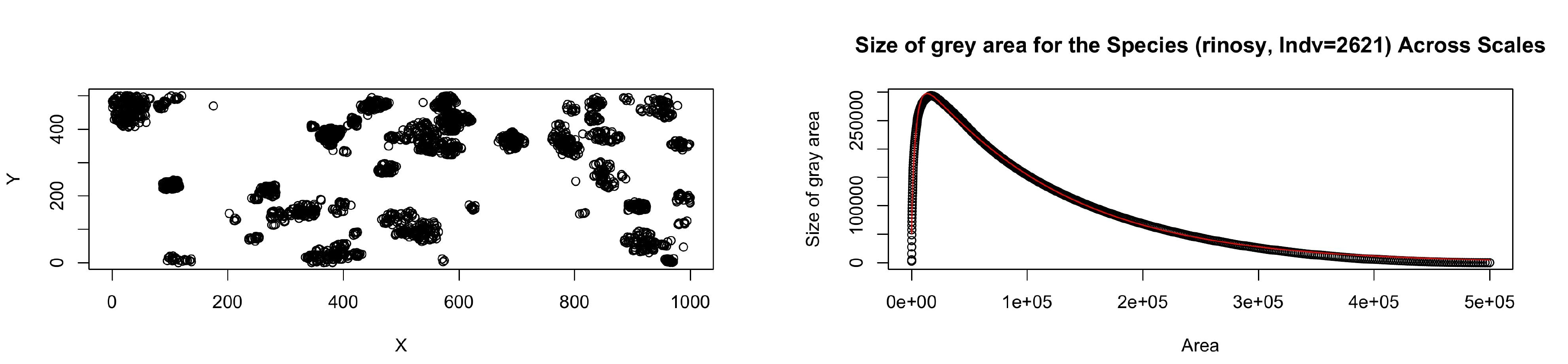}
\includegraphics[width=\linewidth]{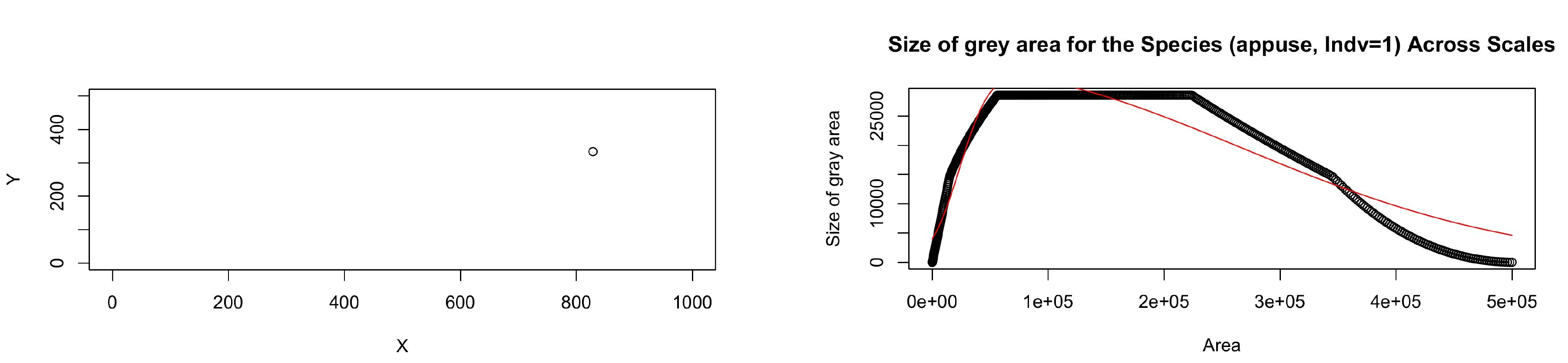}
\caption{For the BCI data, the left figures show how the individuals of the species rinosy and appuse are distributed over the total area of 50 ha, and the right figures show the size of their gray area as a function of the size of the subareas. The red lines are their fits with the parameters in Table \ref{tab2}.}
\label{fig6}
\end{figure}

\begin{figure}[!h]
\centering
\includegraphics[width=\linewidth]{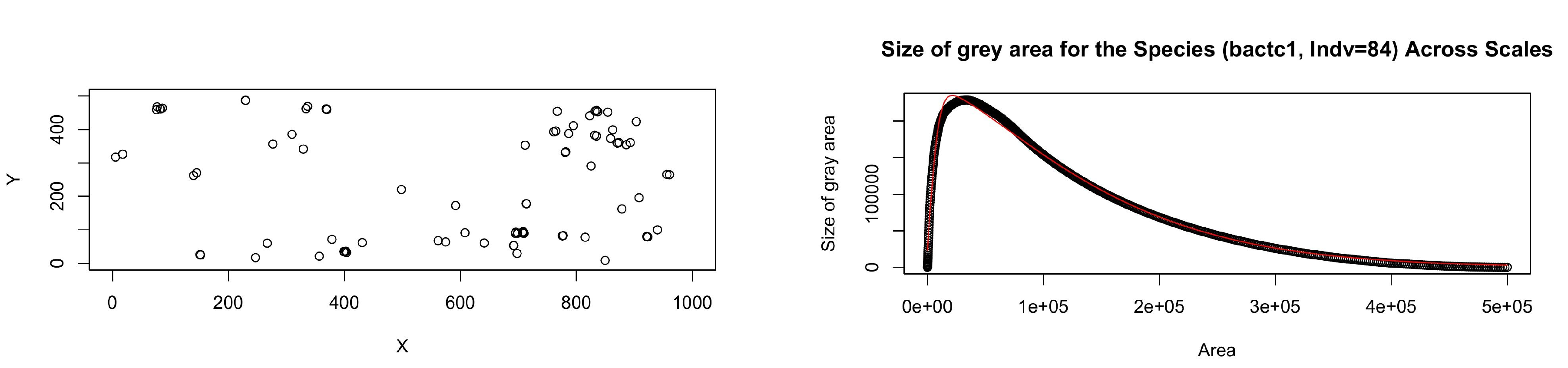}
\includegraphics[width=\linewidth]{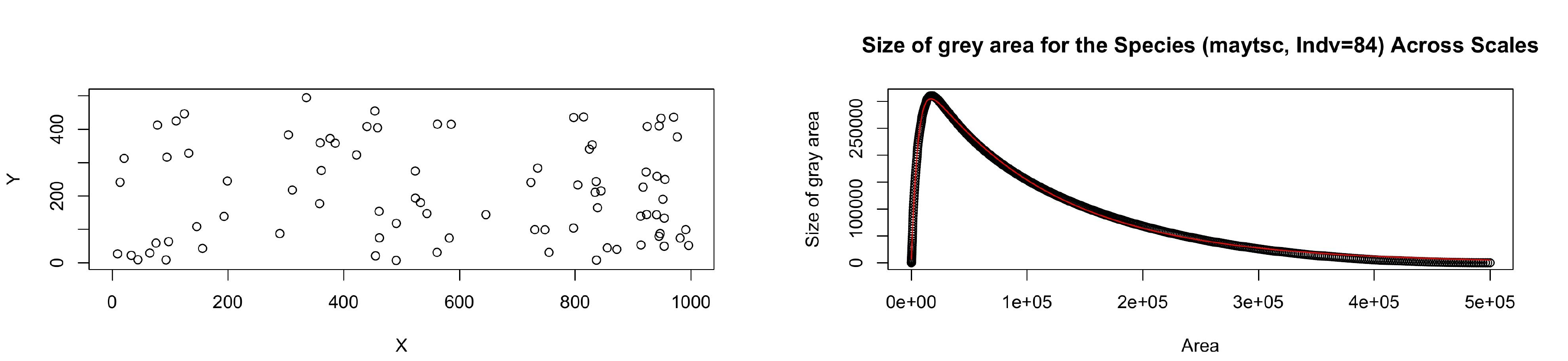}
\caption{For the BCI data, the left figures show how the individuals of the species bactc1 and maytsc are distributed over the total area of 50 ha, and the right figures show the size of their gray area as a function of the size of the subareas. The red lines are their fits with the parameters in Table \ref{tab2}.}
\label{fig7}
\end{figure}

Figures \ref{fig4}, \ref{fig5}, \ref{fig6}, and \ref{fig7} show data from BCI. We plot the positions of individuals of some specific species in the $50$ ha area and plot the size of the grey area as a function of subarea size.


In Figures \ref{fig4}, \ref{fig5}, \ref{fig6}, and \ref{fig7}, we show the size of the gray area across the spatial scale for several species. As we can see, the sharper the maximum peak is, the more widely distributed its individuals are in the area, while being mild means that the individuals are more clustered or it is a rare species. Table \ref{tab2} shows the values of $a_s$, $b_s$, $c_s$, $e_s$ and $\beta_s$ for the species aegipa, anaxpa, cha2sc, entesc, rinosy, appuse, bactc1 and maytsc:

\begin{table*}
\caption{For the BCI data, the table shows the respective values of the parameters for the species maytsc, bactc, appuse, rinosy, entesc, cha2sc, anaxpa, and aegipa. The last column is the value of the chi-square obtained from the formula $\chi^2=\sum\frac{(E-obs)^2}{E}$, where $E$ corresponds to expected values and $Obs$ to observed values.}
\centering
\begin{tabular}{lllllll}
Species Code & $a_s$ & $b_s$ & $c_s$ & $e_s$ & $\beta_s$ & $\chi^2$\\
\midrule
aegipa&0.49&24.76&5.02 e3&6.17&0.41&435.25\\
anaxpa&0.045&15.58&238.97&-51.53&0.21&488.81\\
cha2sc&260.22&1308.72&2.92 e9&37.27&0.92&463.00\\
entesc&2.50&187.12&1.32 e5&3.26&0.54&428.77\\
rinosy&1.35&36.64&4.54 e4&5.93&0.49&440.02\\
appuse&1074.27&23439.47&5.78 e10&11.8&1&442.76\\
bactc1&8.22&213.34&1.95 e6&8.62&0.64&442.07\\
maytsc&0.54&31.88&5.94 e3&5.05&0.42&436.32\\
\bottomrule
\end{tabular}
\label{tab2}
\end{table*}

As we have already explained, the parameter $\beta$ plays the role that enforces the regularization of the dispersal of individuals of the species in the whole area. In order to extract ecological information about the other parameters based on the locations of individuals of a species, we need to force the parameter to be equal to $1$. Table \ref{tab3} is obtained by forcing $\beta_s=1$:
\begin{table*}
\caption{For the BCI data, by omitting the weighting of the subarea sizes, $\beta_s=1$, it shows the respective values of the parameters for the species maytsc, bactc, appuse, rinosy, entesc, cha2sc, anaxpa, and aegipa. The last column is the value of the chi-square. These values also show the significance of the weights of the areas. In front of the values of the parameters we put the standard errors in brackets.}
\centering
\begin{tabular}{llllll}
Species Code & $a_s(SE)$ & $b_s(SE)$ & $c_s(SE)$ & $e_s(SE)$ & $\chi^2$\\
\midrule
aegipa	&558.3 (5.55)	&1907 (64.83)	&9.87 e9 (2.26 e8)	&65.48 (3.99)	&6527.13\\
anaxpa	&2505.8 (164.2)&	420168 (1842)	&7.39 e11 (1.11 e11)	&{\colorbox{green}{-22.91}} (2.65)	&488.70\\
cha2sc	&701.9 (3.08)	&2405 (45.85)	&2.20 e10 (2.24 e8)	&53.82 (1.50)	&488.38\\
entesc	&772.3 (3.39)	&12372 (133.6)	&1.87 e10 (2.07 e8)	&13.83 (0.28)	&519.44\\
rinosy	&568.71 (4.62)	&1065 (49.79)	&1.17 e10 (2.19 e8)	&67.60 (3.95)	&2116.97\\
appuse	&1074.3 (11.26)&	23439 (593)&	5.78 e10 (1.60 e9)&	11.82 (0.54)&	442.76\\
bactc1	&661.6 (3.79)	&3739 (71.17)	&\colorbox{yellow}{1.57 e10} (2.12 e8)	&\colorbox{lightgray}{37.01} (1.18)	&706.92\\
maytsc	&589.7 (5.26)	&2969 (83.99)	&\colorbox{yellow}{1.09 e10} (2.27 e8)	&\colorbox{lightgray}{43.06} (2.16)	&3087.08\\
\bottomrule
\end{tabular}
\label{tab3}
\end{table*}
Pay special attention to the species bactc1 and maytsc, which have the same number of individuals, $84$, but the individuals of bactc1 are more clustered at small distances than the individuals of maytsc, and the individuals of bactc1 appear more on the sides than the individuals of maytsc. Note also that the species anaxpa appears only in the upper left corner of the area.

Now suppose that for a fixed species $s$ the following results (all parameters are preserved):
\begin{equation*}
S(B)\mid_s=\frac{f_s(B)\frac{1}{\sqrt{2\pi c_s}}\exp\left(\frac{-(B^{\beta_s}-b_s)^2}{2c_s}\right)}{(\sqrt{A}-\sqrt{B})^2}.
\end{equation*} 

Fix a species $s$ and consider its individuals as salt in the tank representing the total area. And consider the rest of the individuals as water. We know that the number of individuals as a function of the area follows the linear relationship $N(A)=\rho A$, where $N(A)$ is the number of individuals in area $A$ and $\rho$ is a positive constant. Now let us consider the subarea as time, i.e., for a sequence of subareas $A_1,\ldots,A_k=A$, the constant multiple $\rho$ of the segment sizes $A_{(i+1)}-A_i$ provides the total number of individuals in that segment and is equal to the amount of water moving out of the tank, this amount is considered fixed for all $i$'s, this gives rise to the assumption that $A_{(i+1)}-A_i$ is fixed. Now we are looking for: the probability that at least one individual of species $s$ moves out of the tank as a function of time (area size); according to our previous discussions, for a given subarea size $B$, this probability is $S(B)\mid_s$.


\section{Main Result}
Keeping eyes on the problem of mixed salt-water in a tank, we may formulate the amount of ongoing of the fixed species by the area as follows:
\begin{equation*}
\frac{dS_s}{dB}(B)=S(B)\mid_s,
\end{equation*}

where $\frac{dS_s}{dB}(B)$ means how much of the individuals of species $s$ emigrate (amount of salt from the tank ) when we increase the subarea by $dB$ as a function of the subarea size.

\begin{figure}[!h]
\centering
\includegraphics[width=0.54\linewidth]{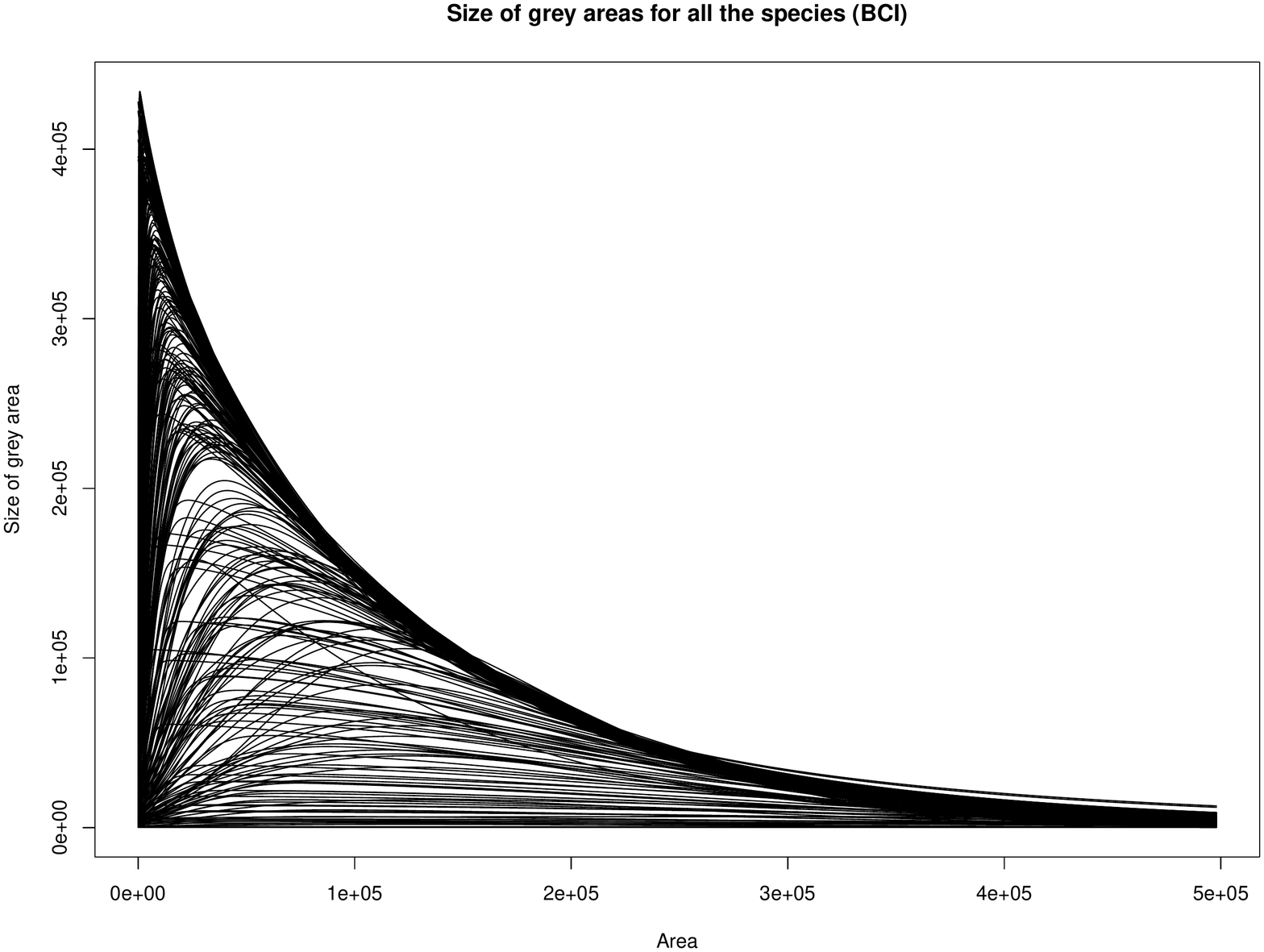}
\includegraphics[width=0.43\linewidth]{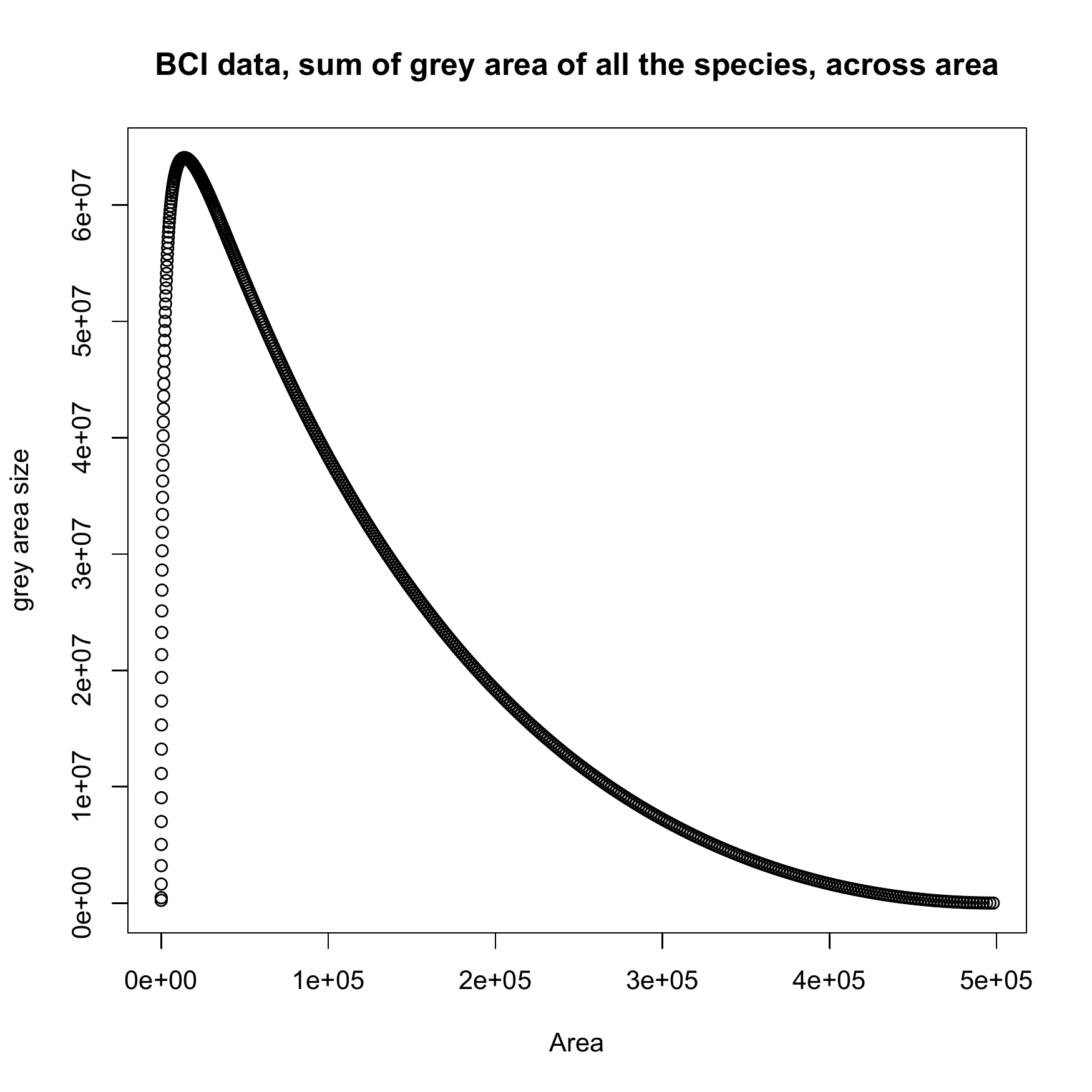}
\caption{For BCI data, the gray areas for all the species and the sum of gray areas of all the species, a similar pattern can be observed.}
\label{fig8}
\end{figure}

Note that the problem of interest is the sums of the previous formulations for all species, where the denominators are fixed for all species, and we only need to add the numerator, which is the respective gray areas for the species. The shape of the sums is in favor of the shape of the gray areas of the common species since their gray areas are larger than those of the rare species. Therefore, we could say that the sum of the gray areas for all species also has a bell shape and so can be handled similarly, see Figure \ref{fig8}.

Then, we have:
\begin{align*}
\frac{dS}{dB}(B)\approx&\frac{\frac{a\delta AS}{\sqrt{2\pi c}}\exp\left(\frac{-(B^{\beta}-b)^2}{2c}\right)}{(\sqrt{A}-\sqrt{B})^2}\times\\
&\left(1+\mathrm{erf}\left(\frac{e(B^{\beta}-b)}{\sqrt{2c}}\right)\right)
\end{align*}
which gives us the total amount of outgoing species by adding $dB$, where $\delta$ is the percentage of the total area A covered by all species. 

Now, if we simplify some of the integrations and use some simplifications with polynomial expressions, we have:
\begin{equation}\label{maineq}
S\approx c'|a'+B^{\beta}|^z,
\end{equation}
where $z=\frac{\delta ap}{Te\sqrt{2\pi c}}$ with $T$ and $p$ are constants obtained from the simplification methods, and $c'$ and $a'$ are constant. A more accurate formulation can be obtained if we use 
$S\approx c'|P_k (B^{\beta})|^{\frac{\delta a p}{Te\sqrt{2\pi c}}},$
where $P_k(B^{\beta})$ is a polynomial of degree $k\geq2$ of $B^{\beta}$, see Appendix \ref{AppB}. 

Note that, the formula \eqref{maineq} is equivalent to say that 
\begin{equation*}
S\approx c'(a'+B^{\beta})^z, B\geq B_0.
\end{equation*}
That is for $S$ agrees with the formula for $B\geq B_0$ where $B_0$ is the smallest subarea size. However, we will keep the original formulation with absolute function because it is the formulation we obtain with our method.

Always keep in mind we want to describe the pattern of SAR, which is the smallest subarea we use. If we change the size of the smallest subarea, the values of the parameters will change accordingly because the adjustment of the sum of the gray areas will change.

\subsection{Applied to Real Data}

In the BCI data, there are some areas without individuals around position $(400,250)$. Likewise, there are some areas around the corner sides without individuals, which explains $\delta\neq 100$. We proceed as follows: First, we solve the differential equation and find the parameters and then compare them with the real data and the Preston power-law fitting, and the second time we find the parameters directly by fitting the real data and again compare them with the real and the power-law fitting. 

Note that, in the BCI data, because log-linear and semi-log-linear are still good fits to describe SAR for the sub-area size from 1 to 50 ha, we will use this interval to perform tests and compare it with our result. We show that in this case, our method is better than both of them. 

\begin{figure}[!h]
\centering
\includegraphics[width=0.6\linewidth]{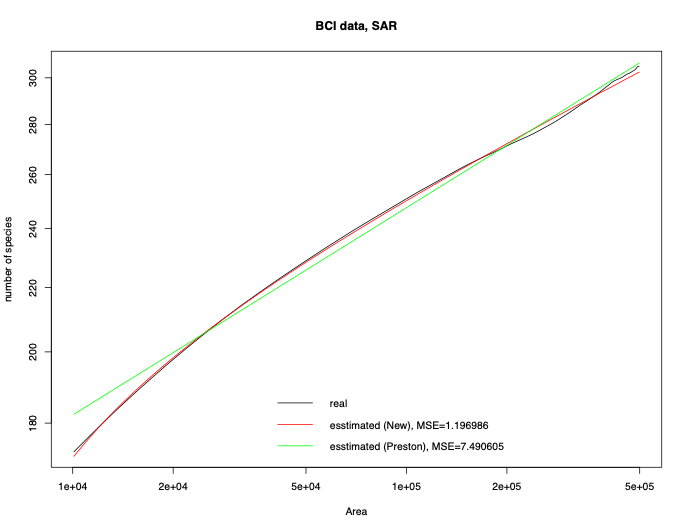}
\caption{Fitting SAR for BCI data, fitting the SAR in a log-log scale and comparing with real SAR by evaluating the mean square error. The black line is the real SAR, the green line is the fitting by power-law, and the red line is the fitting of the new formulations. The area size changes from 1 ha to 50 ha.}
\label{fig9}
\end{figure}

By solving the differential equation and considering the subarea size starting from 1 to 50 ha, we find that $S(B)\approx30.36|33721.11-37.76B^{0.6} |^{0.17}$ (see Figures \ref{fig9}). Note that, if we consider the smallest subarea to be larger, the fit of the gray areas changes accordingly\footnote{This is similar to the proposed veil line of \cite{Preston1948}, we could see that by increasing the size of the smallest subarea, the line is shifted, but now to the right side.}. As we have already described, the parameters of the simplified formula \eqref{maineq} depend on the locations of the individuals of the species and the size of the smallest segment of the area. The formula \eqref{maineq} is obtained by general fitting the size of the gray areas. The gray areas corresponding to individuals of a species may overlap depending on the clustering of species. Therefore, all extrapolations for smaller subareas must take into account the clustering of species in smaller subareas, see Appendix \ref{AppC}.

\begin{figure}[!h]
\centering
\includegraphics[width=0.6\linewidth]{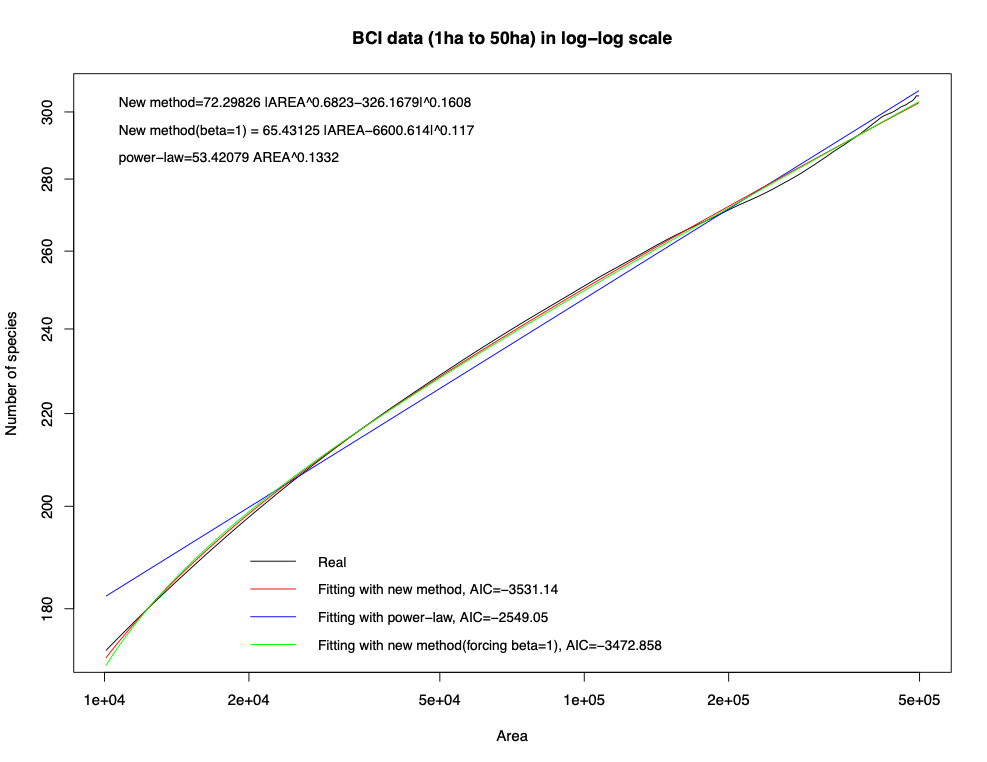}
\caption{Fitting SAR for BCI data, fitting the SAR on a log-log scale and comparing it to the real SAR by evaluating the respective AIC's. The black line is the real SAR, the blue line is the fitting by power-law (Preston), the red line is the fitting of the new formulas $\beta\neq1$, and the green line is the new formula with $\beta =1$. The area size changes from 1 ha to 50 ha.}
\label{fig10}
\end{figure}

Now we fit the data with the formula \eqref{maineq}, we also consider the case where $\beta=1$ and compare it with the real data and the power- law, see Figure \ref{fig10}.

Figure \ref{fig10} shows that the fit using our method has a smaller difference to the number of species in the 1 ha subarea than that using the power-law. Since our model is a downward convex function, but the power-law is flat at the log-log scale, extrapolation to a valid smaller scale (no extreme cases) shows that our result favors the exponential decay in species number without running any tests.


\subsection{Fitting with Beta and the Gamma Distributions}\label{gama}

Here we have examined the possibility of fitting the sum of gray areas with Beta and Gamma distributions (because of their properties and their shapes). We applied the fitting of a gray area as a multiple of Beta and Gamma distributions and considering the boundary conditions.

Let us first consider the fitting of the sum of gray areas with a constant multiplier of  Beta distribution:
\begin{equation}\label{bet}
B(t,,\alpha,\beta,c)=\frac{ct^{\alpha-1}(1-t)^{\beta-1}}{\mathrm{Beta}(\alpha,\beta)}.
\end{equation}
We have assumed that the sample area is rectangular, and the ratio of height to width is $a$. Hence, the size of the subarea with the width equal to $y$ is $ay^2$. To fit with Beta, we need to convert the size of the subarea to the closed interval of $0$ and $1$. To do this, we just need to transform the size of the area with the following transformation:
\begin{equation}\label{trnsfrm}
T:x\rightarrow\frac{\sqrt{\frac{x}{a}}}{X},
\end{equation}
where $X$ is the height of the total area of width $Y$. If we assume that the size of the subarea is equal to $x$, then the transformation $t=T(x)$ is used in the formula \eqref{bet}, which implies $x=a(Xt)^2$. Now $\frac{B(t,\alpha,\beta,c)}{(\sqrt{A}-\sqrt{x})^2}$ describes the SAR across scale, which implies

\begin{equation*}
\frac{\frac{ct^{\alpha-1}(1-t)^{\beta-1}}{\mathrm{Beta}(\alpha,\beta)}}{A(1-t)^2}=\frac{ct^{\alpha-1}(1-t)^{\beta-3}}{A\mathrm{Beta}(\alpha,\beta)}.
\end{equation*}
Note that $x$ tends towards $A$, which means that $1-t$ tends towards $0$. So for the above function to make sense, we must have $(1-t)^{\beta-3}=1$ when $t=1$, which means $\beta=3$. So, we have
\begin{align*}
S(x)&=\frac{cT(x)^{\alpha-1}}{A\mathrm{Beta}(\alpha,3)}\\
&=\frac{c\left(\sqrt{\frac{x}{A}}\right)^{\alpha-1}}{A\mathrm{Beta}(\alpha,3)}\\
&=\frac{cA^{\frac{\alpha-3}{2}}}{\mathrm{Beta}(\alpha,3)}x^{\frac{\alpha-1}{2}},
\end{align*}
which is simply a power law. In the case of the BCI data, we get $\alpha=1.2664$ and $c=32711956$. 

If we ignore the boundary condition and assume $\beta\neq3$ and start fitting directly, then the result for different sample data will be invalid for either a small subarea or a large subarea close to the total area.

If we now consider the Gamma distribution, we have:

\begin{equation*}
S(x)=\frac{G(t,\alpha,\beta,c)}{(\sqrt{A}-\sqrt{x})^2},
\end{equation*}
where $G(t,\alpha,\beta,c)=c\frac{\beta^{\alpha}}{\Gamma(\alpha)}t^{\alpha-1}e^{-\beta t}$ and $t$ is the width of sub-area $x$. For $a$ equal to the proportion of height by width we have $(\sqrt{A}-\sqrt{x})^2=a(X-t)^2$, where $X$ is the width of the total area, and $t=\sqrt{\frac{x}{a}}$. This implies 
\begin{equation*}
S(x)=\frac{c}{a}\frac{\beta^{\alpha}}{\Gamma(\alpha)}\frac{t^{\alpha-1}e^{-\beta t}}{(X-t)^2}.\end{equation*}
Since $S(A)$ is the total number of species in the total area $A$, the term $\frac{t^{\alpha-1}e^{-\beta t}}{(X-t)^2}$ must have meaning while $t\rightarrow X$. Let $u=1-\frac{t}{X}$, then $t\rightarrow X$ implies $u\rightarrow0$. By interchanging the variable, we have the following term:
\begin{equation*}
\frac{c}{a}\frac{\beta^{\alpha}}{\Gamma(\alpha)}X^{\alpha-3}\frac{(1-u)^{\alpha-1}e^{-\beta X(1-u)}}{u^2}.\end{equation*}
When $u\rightarrow 0$, the denominator must be cancelled, which implies the numerator for $u\rightarrow0$ must be equal to $0$. If $\alpha$ and $\beta$ are bounded, then invalid result will appear according to the fact that $u\rightarrow0$ implies the infinity in the final result. So,  at least one of $\beta$ and $\alpha$ are infinite. However, considering a Gamma distribution under the assumption that at least one of the parameters are infinite is invalid for fitting data, which means that the use of Gamma distribution is not proper, but we will explore these cases for clarification. In case assuming both are infinite or $\beta=\infty$ and $\alpha$ is finite, then, with re-arrangement, the term
\begin{equation*}
\frac{c}{a}\frac{1}{\Gamma(\alpha)}X^{\alpha-3}\frac{(1-u)^{\alpha-1}e^{-\beta X(1-u)}}{\beta^{-\alpha}u^2}\end{equation*}
is $0$ almost everywhere. So, we can only consider $\alpha=\infty$ and $\beta$ to be finite. So, we have the following re-arrangement term:
\begin{equation*}
\frac{c}{a}\frac{X^{\alpha-3}}{\Gamma(\alpha)}\frac{1}{\beta^{-\alpha}}(1-u)^{\alpha-1}\frac{1}{u^2}e^{-\beta X(1-u)}.
\end{equation*}
By assumption $u\rightarrow0$, the terms $\frac{X^{\alpha-3}}{\Gamma(\alpha)}$ and $(1-u)^{\alpha-1}$ tends to zero (with multiplicity\footnote{Assume that $f(x)=(x-a)^k g(x)$ such that $g(a)\neq0$, then $k$ is called the multiplicity of $(x-a)$ in $f(x)$. In other words, the maximum number of zero factors in a term} 1), which suffices to cancel the term $u^2$. Now, it is easy to see that the term
\begin{equation*}
\frac{c}{a}\frac{\beta^{\alpha}}{\Gamma(\alpha)}X^{\alpha-3}\frac{(1-u)^{\alpha-1}e^{-\beta X(1-u)}}{u^2}\end{equation*}
is zero if $\beta<1$ and is infinity if $\beta>1$. So,  we must have $\beta=1$. Hence, 
\begin{equation*}
S(x)=\frac{c}{a}\frac{t^{\alpha-1}}{\Gamma(\alpha)}\frac{e^{-t}}{(X-t)^2},\end{equation*}
where $\alpha$ is very large and $t=\sqrt{\frac{x}{a}}$.

\begin{figure}[!h]
\centering
\includegraphics[width=0.6\linewidth]{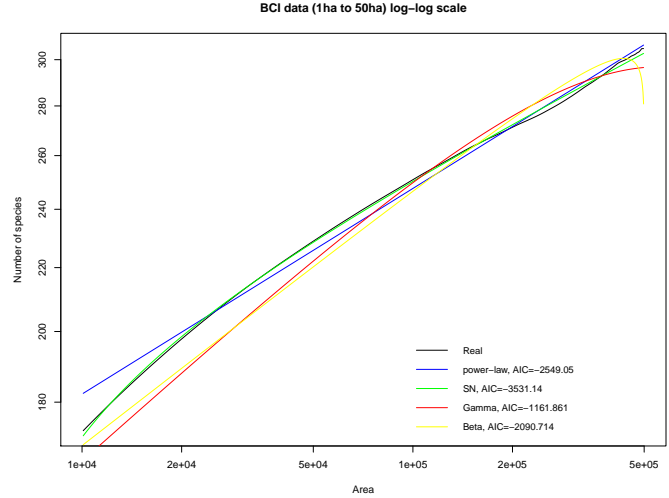}
\caption{Fitting SAR for BCI data, fitting the SAR on a log-log scale, and comparing it to the real SAR by evaluating the respective AIC's. The black line is the real SAR. The blue line is by power-law, the red line by Gamma, the green line by the new formula, and the yellow line by Beta. The area size changes from 1 ha to 50 ha.}
\label{fig11}
\end{figure}

We apply the fit with Beta and Gamma distribution to the BCI data, Figures \ref{fig11}, without considering the boundary conditions. We avoid the plot with the boundary conditions since the Beta distribution implies the power- law and the Gamma distribution implies infinity as a parameter.

\section{Discussion}

The log-log linear or power-law \cite{Preston1962} and semi-log linear \cite{Gleason1922} patterns for SAR can describe SAR only for large areas, but both have difficulty for small areas. The formula as a sum of polynomials \cite{Conceicao2014} does not describe the correlation between coefficients; this may lead to overfitting. Moreover,  for a very small area, they involve a computational error due to the existence of logarithm and inverse of a very small area. We propose that the pattern of SAR follows a log-log power law instead of the log-log linear one, which explains the curvature of the pattern of SAR for a small area, Figure \ref{fig9}. The new formulation is also consistent with the relationship of the exponent (the external exponent in our formula) as the environmental component, as we explained in the content the relationship with spatial aggregation, percent area coverage, and the core-satellite model. We suggested that SAR can be formulated as $S=c|A^{\beta}+a|^z$, which can be considered as a power law on the log-log scale or the semi-logarithmic scale of the area, while the formulations in previous works on SAR suggest that SAR is linear on the semi-logarithmic scale of area or the log-log scale. However, as we can see in Figure \ref{fig9}, the straight line, in fact, occurs only for large areas. It is easy to see that for a large area, the value of $a$ is negligible, which induces a log-log linear and a semi-log linear function. This implies that the formula for large areas can be interpreted as a power-law or a semi-log linear function. Therefore, our formula can be considered as a generalized version.

Note that in our formulation, if we agree to go through the solution of the differential equation, all parameters $a$, $\beta$, and $z$ can be obtained from the estimates, and the only parameter we need to fit with the data is $c$. For a sufficiently large area, SAR can be formulated as $S\approx ca^z A^{\beta z}$, which is the log-log linear formulation of \cite{Preston1962} where all available parameters can be obtained from the data. Also, if we omit the estimation with $p$ as the numerical estimator of the exponential in Appendix \ref{AppB} and instead use the polynomial expansion of the exponential (this part is in Remark \ref{C1}) and together with omitting the corresponding area weight $\beta=1$, then the result of \cite{Conceicao2014} can be obtained. In the technique invented by \cite{Conceicao2014}, they adjust the coefficients of the powers of the logarithm of the area, the area, and the inverse of the area. But in our result, depending on the degree of accuracy, not only the values of the coefficients can be found, but also the formulas for obtaining the coefficients from the parameters can be easily found. Moreover, all formulas \cite{Preston1962, Gleason1922, Conceicao2014} can be extracted from our result.

The parameter $\beta$ represents a weighting for the area size, related to the differences in the rate of increase and decrease of the sums of the gray areas for all species. We do not yet know whether the value of $\beta$ has any ecological significance.

In our case, we use a simple estimation for the $\mathrm{erf}$-function, which is a polynomial of degree $2$. An interesting approach would be to estimate $\mathrm{erf}$ by polynomials with degrees larger than $2$. Another interesting approach could be to use a recursive formula. In this case, we need to know the number of species for a very small area. Then use a recursion algorithm to find the number of species for a larger area. This approach has a very high complexity. See Appendix \ref{AppA} that we conjecture a possible formulation.

\begin{figure}[!h]
\centering
\includegraphics[width=0.6\linewidth]{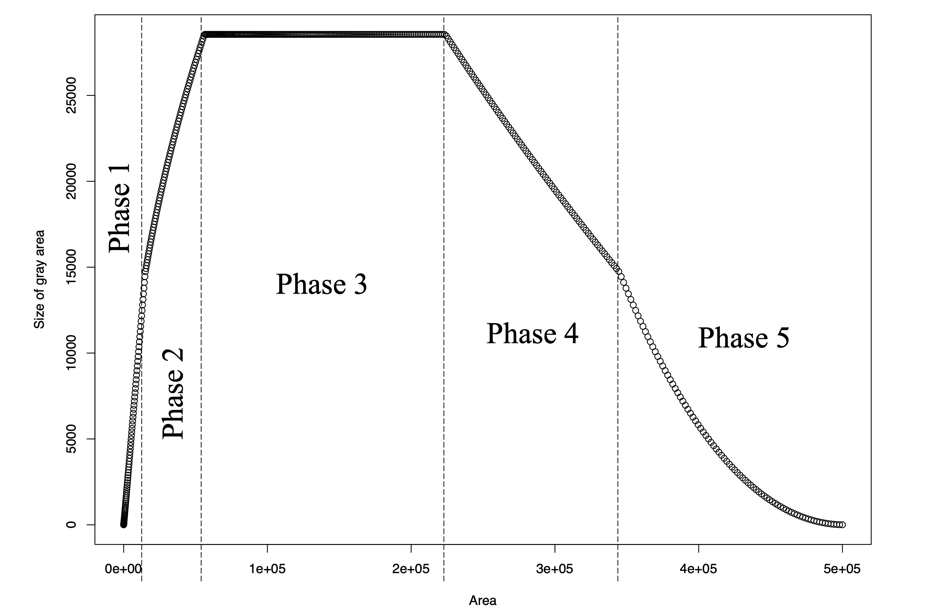}
\caption{Shows one way (the extreme scenario) of identifying different phases for a plot of the gray area.}
\label{figl}
\end{figure}

As another approach, we can call it weaker because of the fit with simple functions, and we can consider some partial functions to describe the gray areas in different phases. Each phase depends on the non-differentiable points on the plot of the gray areas. The extreme case can be seen in Figure \ref{figl}, where for the plot of the change of a gray area we have $5$ phases, the first $4$ being linear as a function of the area size. The changes in the phases correspond to the changes in the gray areas relative to the changes in the black and dashed areas. And the choice of rational functions are the main difficulties of this approach; see Appendix \ref{AppA} for a possible formulation.

However, we believe that the previous approach gives a similar result in a simplified version. The difference is in our approach; the relationship with the environmental components can be better explained.

\subsection{Conclusion}

Our result is based on the positions of the individuals of the species in the area, and we used some approximations to simplify the result. When it comes to accuracy, we can improve the approximations in a way that is explained in the content. In our method, we used the most simplified formulas and approximations, but the mean squared errors are lower than the other results; this means that our result, even using some weak estimates, better explains the pattern of SAR. Another advantage of the new method is that it allows us to constrain a single species, and it better explains SAR for small area sizes.

The main limitation of our studies is the use of more terms to improve the fit. Additional terms introduce complexities in the integrations (see Appendix \ref{AppB}), making it harder to find a simple formulation. The $\mathrm{erf}$-function that occurs in the content does not allow us to do a simple integration, so we used a simple estimate for it, which is a polynomial of degree $2$. If we use a different estimate, then the complexity of the integration grows.

\section{Acknowledgments}
This research was partially supported by Funda\c{c}\~{a}o para a Ci\^{e}ncia e a Tecnologia (PTDC/BIA-BIC/5558/2014) as the BPD project species diversity as a function of spatial scales with reference number ICETA-2016-43. The BCI forest dynamics research project was founded by S.P. Hubbell and R.B. Foster and is now managed by R. Condit, S. Lao, and R. Perez under the Center for Tropical Forest Science and the Smithsonian Tropical Research in Panama. Numerous organizations have provided funding, principally the U.S. National Science Foundation, and hundreds of field workers have contributed. We also acknowledge R. Foster as plot founder and the first botanist able to identify so many trees in a diverse forest; R. P\'{e}rez and S. Aguilar for species identification; S. Lao for data management; S. Dolins for database design; plus hundreds of field workers for the census work, now over 2 million tree measurements; the National Science Foundation, Smithsonian Tropical Research Institute, and MacArthur Foundation for the bulk of the financial support.


\bibliographystyle{apalike}
\bibliography{all}
\appendix
\section{Tools and Ideas}\label{AppA}
\counterwithin{figure}{section}
\subsection{Obtaining SAR for a Given Sample Data}
Let a sample consist of a fixed area size $A$, species $1,\ldots, S$, and positions of their individuals in the area. Consider the following nested sequence of subareas of $A$,
\begin{equation*}
P:0\neq A_1\subset\cdots\subset A_n=A,
\end{equation*}
where these subareas are randomly chosen. Then we have
\begin{equation*}
S_P:0\neq S_p(A_1)\leq\cdots\leq S_P(A_n)=S,
\end{equation*}
where $S_P (A_i)$ is the number of species in the subarea $A_i$. Now randomly choose $m$ sequences that are similar to $P$ in the sense that for each such sequence 
\begin{equation*}
P_j:0\neq B_{1,j}\subset\cdots\subset B_{n,j}=A,
\end{equation*}
where $j=1,\ldots,m$, and for each $i=1,\ldots,n$, the equalities $|A_i |=|B_{i,j} |$ hold, where $|A_i |$ is the size of $A_i$. We slightly abuse the notation and denote both $|A_i |$ and $A_i$ by $A_i$. 

Let $B$ be a subarea of $A$ with size equal to $A_i$ and let $S(B)$ be the mean of the number of species of the area of the $i$-th position of all $m$ sequences
\begin{equation*}
S(B)=\frac{\sum_{j=1}^mS_{P_j}(B_{i,j})}{m}.
\end{equation*}
Since the number of species is bounded by $0$ and $S$, $S(B)$ converges for sufficiently large $m$. So we can define $S(B)$ for any subarea $B$ of $A$, which allows us to find SAR. That $S(B)$ is well-defined, see \cite{Willard2004}.

On the other hand, since the nested sequences are chosen randomly, we can have the following restriction over $S_P (A_i )$ for any sequence $P$ and for a fixed species $s$:
\[
S_P(A_i)\mid_s=\left\{
\begin{array}{lll}
 0 &   & s\notin A_i  \\
  &   &   \\
 1 &   & s\in A_i  
\end{array}
\right..
\]
It follows, 
\begin{equation*}
S_P (A_i )=S_P (A_i ) \mid_1+\cdots+S_P (A_i )\mid_S.
\end{equation*}
This means that for a given subarea with a random position in the total area, we can somehow consider its species number as a binary code corresponding to the present and absent species in that subarea. This is of great use in our approach as it is related to the probability of observing a species.

We can translate $S(B) \mid_s$ as the probability of observing species $s$ in the subarea of size $B$. And
\begin{equation*}
S(B)=S(B) \mid_1+\cdots+S(B) \mid_S,
\end{equation*}
where 
\begin{align*}
S(B) \mid_s&=\mathrm{mean}(S_{P_j } (B) \mid_s , j=1,\ldots,\infty)\\
&=\lim_{m\rightarrow\infty}\frac{\sum_{j=1}^{m}S_{P_j}(B)\mid_s}{m}.
\end{align*}

\subsection{Geometric Approach}
The data of a rectangular sample would consist of a fixed area size $A$, the species $1,\ldots, S$, and the positions of their individuals in the area.
\begin{remark}
In this paper, by a subarea, we mean a subarea where the ratio of length to width is equal to the ratio of length to width of the total area.
\end{remark}
\begin{lemma}\label{Lem}
Each subarea of size B can be identified by a point that is the point of the upper left corner of that subarea.\qed
\end{lemma}
In other words: If we remove the band of width equal to the width of $B$ and the band of length equal to the length of $B$ from the right and bottom areas, respectively, then there is a one-to-one correspondence between subareas of size $B$ and points in the remaining area. All points in the two bands on the right and bottom of the total area cannot represent a subarea because of the width and length.

For a fixed species $s$ and a fixed size as the size of a subarea $B$, Lemma \ref{Lem} allows us to partition the total area into three distinct subareas, the first partition containing only the two bands in the previous paragraph to which we cannot assign a subarea, and another partition, having a one-to-one correspondence to the subareas of size $B$, this partition can be considered by itself as two partitions: One containing all the points where there is at least one individual of species $s$ in the respective subarea of size $B$, and another containing the other points. In other words, there are three partitions:
\begin{itemize}
\item The area where we cannot match a subarea of size $B$;
\item The area in which we can match a subarea of size $B$ and the corresponding subareas contain at least one individual of species $s$;
\item The area in which we can match a subarea of size $B$ and the corresponding subareas contain none of the individuals of species $s$.
\end{itemize}
The second and third partitions suggest that we can assign to each of the individuals of species $s$ an area that we call the gray area, see Figure \ref{fig3}, and it is the largest rectangular area of size at most $B$ within the total area where the position of the individual is in the lower right corner of this area. We call the third partition the blue dashed area and the first the black dashed area.

\subsection{Recursive formulation}

Considering the recursive formulation, a simplified formulation should have the following form:
\begin{equation*}
S(A+\delta)=\alpha S(aA^{\beta})+C,
\end{equation*}
where $\alpha$, $C$, $\beta$, and $\delta$ are constants, and $\delta$ denotes the increasing area intercept. The initial value is: 
\begin{equation*}
S(A_0 )=S_0.\end{equation*}
This formulation simply means: 
\begin{equation*}
S(A)=c(aA^{\beta}-A'_0 )^{\alpha},
\end{equation*}
for some constants $\alpha,c,\beta$, and $A'_0$.

\subsection{Consideration of $5$-phases for changing the size of gray areas}
By considering $5$-phases as in Figure \ref{figl}, a simplified formulation should read:
\[
\frac{dS}{dB}=\frac{\left\{
\begin{array}{lll}
 a_1B^{\beta}+b_1 & & ,\text{phase} 1\\
 a_2B^{\beta}+b_2  & & ,\text{phase} 2\\
 a_3B^{\beta}+b_3  & & ,\text{phase} 3\\
 a_4B^{\beta}+b_4  & & ,\text{phase} 4\\
 \frac{a_5}{B^{\beta}+b_5} & & ,\text{phase} 5
\end{array}
\right.}{(\sqrt{A}-\sqrt{B})^2}.,
\]
where $a_i$ and $b_i$ are constants, $ \frac{a_5}{B^{\beta}+b_5}$ is the simplest rational function fitted to the data, $\beta$ is a constant corresponding to the area weights, and the result can be obtained by integrations. A better estimator than a rational function for phase $5$ gives a better result. 

\subsection{Why 5 Parameters}
Looking at Figure \ref{figl}, it is easy to see that to fully describe the gray area, at least $3$ parameters are needed, the slopes of phases 1,2 and 4, and at least one parameter must be added to describe phase 5, and one parameter must be added as random noise. Therefore at least $5$ parameters are needed to describe the pattern of the gray area. 

In mathematical terms, if we want to estimate a good fit for a given figure of a continuous function without prior information, we need a corresponding parameter in the fit for each of the following points in the figure. 1) roots 2) local max and min values 3) the points at which the figure changes convexity. In doing so, we will not consider any correlation between all the preceding that may occur. But if we do consider correlation, then we have a reduction to one parameter for each. For example, if a normal distribution figure is given, then we need 5 parameters without considering correlation. But then 2 roots appear at infinity, and the 2 convexity change points are mirrored, so we can drop the first 2 parameters for roots and consider only 1 for convexity for the last 2. If we know the position of the maximum point, then only 1 parameter is sufficient.

\subsection {Estimation by Rational Expressions }
Recall the following estimations by rational expressions. These can be found in \cite{Abramowitz1964}, the first estimate is from \cite{Heald1985}
\begin{enumerate}
\item $\mathrm{erf}(x)\approx1-\frac{1+0.506x}{\sqrt{2}+2.054x+1.79x^2};$
\item $\exp(x)\approx1+x+\frac{x^2}{2};$
\item for $x<1$:
\begin{itemize}
\item $\frac{1}{1-x}=1-x+x^2$;
\item $\frac{1}{(1-x)^2}=-1+2x-3x^2$.
\end{itemize}
\end{enumerate}

\section{Estimation for Solution of the Differential Equation}\label{AppB}
We assume that species are not partitioned in the sense that species are only classified as either very rare or very abundant. This is the case when the species abundance distribution has the $U$-shape. 

Recall that the sum of gray areas is in favor of the species with the greatest area coverage, in the sense that they quickly reach their maximum value of gray areas, and this value is higher with respect to other species so that in the aggregate, they have a greater advantage than other rare or very abundant species. Therefore, the sum of gray areas follows a similar bell-shaped pattern as a single species.

Recall that for a species $s$, $S(B)\mid_s=\frac{\varepsilon_s(B)}{(\sqrt{A}-\sqrt{B})^2}$. And the proper formulation is:
\begin{equation*}
S(B)\mid_s=\frac{f_s(B)\frac{1}{\sqrt{2\pi c_s}}\exp\left(\frac{-(B^{\beta_s}-b_s)^2}{2c_s}\right)}{(\sqrt{A}-\sqrt{B})^2}
\end{equation*}
where 
\begin{equation*}
f_s(B)=\alpha_s\left(1+\mathrm{erf}(\frac{e_s(B^{\beta_s}-b_s)}{\sqrt{2c_s}})\right).
\end{equation*}

Recall also that 
\begin{equation*}
\frac{dS}{dB}=\sum_iS(B)\mid_i=\sum_i\frac{\varepsilon_i(B)}{(\sqrt{A}-\sqrt{B})^2},
\end{equation*}
where the numerator is the sum of the gray areas. Now, we can factorize a bell-shaped pattern $\varepsilon(B)$ from the numerator, which means:
\begin{equation*}
\frac{dS}{dB}=\frac{\varepsilon(B)\sum_i\frac{\varepsilon_i(B)}{\varepsilon(B)}}{(\sqrt{A}-\sqrt{B})^2}.
\end{equation*}
We have $\sum_i\frac{\varepsilon_i(B)}{\varepsilon(B)}=\delta AS$, because the sum is the percentage of the total area $A$ covered by all species. By replacing $\varepsilon(B)$ with the following term

\begin{align*}
&\frac{\alpha}{\sqrt{2\pi c}}\exp\left(\frac{-(B^{\beta}-b)^2}{2c}\right)\\
&\times\left(1+\mathrm{erf}\left(\frac{e(B^{\beta}-b)}{\sqrt{2c}}\right)\right)
\end{align*}
and with a transformation we have
\begin{align*}
\frac{dS}{dB}\approx&\delta AS \alpha\\
&\times\frac{\exp\left(\frac{-(B^{\beta}-b)^2}{2c}\right)\left(1+\mathrm{erf}\left(\frac{e(B^{\beta}-b)}{\sqrt{2c}}\right)\right)}{\sqrt{2\pi c}(\sqrt{A}-\sqrt{B})^2}.
\end{align*}
Now, we proceed to solve the differential equation. By interchanging with the rational expression in Appendix \ref{AppA}:
\[\displaystyle{
\begin{array}{l}
\ln(S)\approx\int\frac{\delta\alpha}{\sqrt{2\pi c}}\times\\
\left(2-\frac{1+0.506\frac{e(B^{\beta}-b)}{\sqrt{2c}}}{\sqrt{2}+2.0541\frac{e(B^{\beta}-b)}{\sqrt{2c}}+1.79\left(\frac{e(B^{\beta}-b)}{\sqrt{2c}}\right)^2}\right)\\
\times\left(1-\frac{(B^{\beta}-b)^2}{2c}\right)\left(-1+2\sqrt{\frac{B}{A}}-3\frac{B}{A}\right)dB.\end{array}}
\]
Recall that, we are interested in the description of the pattern for small $B$. However, the arguments we used works for sufficiently large and sufficiently small $B$. In these cases, the integration term can be estimated as:
\[\displaystyle{
\begin{array}{l}
\ln(S)\approx\frac{-\delta\alpha}{\sqrt{2\pi c}}p\times\\
\int\left(2-\frac{1+0.506\frac{e(B^{\beta}-b)}{\sqrt{2c}}}{\sqrt{2}+2.0541\frac{e(B^{\beta}-b)}{\sqrt{2c}}+1.79\left(\frac{e(B^{\beta}-b)}{\sqrt{2c}}\right)^2}\right)dB,\end{array}}
\]
where $p> 0$ is a single numerical estimate for the constrained function $\exp\left(\frac{-(B^{\beta}-b)^2}{2c}\right)$ depending on whether the value of $B$ is sufficiently large or small as the upper or lower bound of the function.
\begin{remark}\label{C1}
To achieve the result of \cite{Conceicao2014}, in this step instead of $p$, we need to insert a polynomial representing $\exp\left(\frac{-(B^{\beta}-b)^2}{2c}\right)$ and we also need to drop the weight to the area that is $\beta=1$. Then the integration term is the division of two polynomials, whose solution can be simplified. If the solution is a $\tan^{-1}$ function, then substitution with a polynomial representative is necessary \cite{Abramowitz1964}. 
\end{remark}
Now, back to the solution of the integration. Interchanging the variables $D=\frac{(B^{\beta}-b)}{\sqrt{2c}}$, we get $dD=\frac{\beta B^{\beta-1}}{\sqrt{2c}}dB$, which implies $dB=(D\sqrt{2c}+b)^{\frac{1-\beta}{\beta}}dD$. Thus, by rearranging the terms, we obtain

\[\displaystyle{
\begin{array}{l}
\ln(S)\approx\frac{\delta\alpha}{\beta\sqrt{\pi}}p\times\\
\int\left(\frac{(1+0.506eD)(D\sqrt{2c}+b)^{\frac{1-\beta}{\beta}}}{\sqrt{2}+2.0541eD+1.79\left(eD\right)^2}\right)dD\\
-\frac{\delta\alpha}{\sqrt{2\pi c}}pB+\text{constant}.
\end{array}}
\]
Since $0<\beta< 1$, there is an integer $k\geq0$, such that
\[\displaystyle{
\begin{array}{l}
\ln(S)\approx\frac{\delta\alpha}{\beta\sqrt{\pi}}p\times\\
\int\left(\frac{(1+0.506eD)(D\sqrt{2c}+b)^k}{\sqrt{2}+2.0541eD+1.79\left(eD\right)^2}\right)dD\\
-\frac{\delta\alpha}{\sqrt{2\pi c}}pB+\text{constant}.
\end{array}}
\]
Then we perform polynomial division in the integration term. The following integration is bounded, for some values $T_1$ and $T_2$, and also the constants $G_1$ and $G_2$ (can be obtained from the remainder of the polynomial division). The bound is obtained by increasing or decreasing the coefficients of the denominator to find a perfect logarithmic form in the integration:
\[\displaystyle{
\begin{array}{l}
\frac{1}{eT_2}\ln\left(\sqrt{2}+G_1eD+\frac{G_2}{2}(eD)^2\right)\leq\\
\int\left(\frac{(G_1e+G_2eD)}{\sqrt{2}+2.0541eD+1.79\left(eD\right)^2}\right)dD\leq\\
\frac{1}{eT_1}\ln\left(\sqrt{2}+G_1eD+\frac{G_2}{2}(eD)^2\right).
\end{array}}
\]
Thus, by the result of polynomial division, the following general formula is obtained:
\begin{equation*}
\ln(S)\approx\frac{\delta\alpha p}{Te\sqrt{2\pi c}}\left(\ln(P_2(eD))+P_k(eD)\right)+C,
\end{equation*}
where $P_k(eD)$ represents a polynomial of degree $k$ of variable $eD$ which is the quotient of the polynomial division $\left(\frac{(1+0.506eD)(D\sqrt{2c}+b)^k}{\sqrt{2}+2.0541eD+1.79\left(eD\right)^2}\right)$ and 
\begin{equation*}
P_2(eD)=\sqrt{2}+G_1eD+\frac{G_2}{2}(eD)^2,\end{equation*}

and $0<\delta\leq100$, $p> 0$, and $T$ is from the previous bound $T_1\leq T\leq T_2$ (for simplicity we can choose $T=\frac{T_1+T_2}{2}$) which respects the remainder of polynomial division. Thus, by interchanging the variables $D$, we have:
\begin{equation*}
\ln(S)\approx\frac{\delta\alpha p}{ TE \sqrt{2\pi c}}\left(\ln(P_2(B^{\beta})\exp(P_k(B^{\beta}))\right)+C.
\end{equation*}
Now, we can consider the polynomial expression of the exponential function:
\[\displaystyle{
\begin{array}{l}
\ln(S)\approx C+\frac{\delta\alpha p}{ TE \sqrt{2\pi c}}\times\\
\ln\left(P_2(B^{\beta})\left(1+P_k(B^{\beta})+\frac{(P_k(B^{\beta})^2}{2!}+\cdots\right)\right).
\end{array}}
\]
It follows that,
\begin{equation*}
S\approx c'|P_n(B^{\beta})|^{\frac{\delta\alpha p}{ TE \sqrt{2\pi c}}},
\end{equation*}
where $P_n$ represents a polynomial of degree $n$. Now, if we omit the terms with degree higher than $\beta$:
\begin{equation*}
S\approx c'|a_1+a_2B^{\beta}|^{\frac{\delta\alpha p}{ TE \sqrt{2\pi c}}}.
\end{equation*}
So by reformulation, we have
\begin{equation*}
S=c|B^{\beta}+a|^z.
\end{equation*}

\section{Clustering}\label{AppC}
The world is organized by classifications, as witnesses look at species in biology, diseases in medicine, continents in geography, parties in politics, sets in mathematics, and so on. One of the ways to classify individuals of a species in a given area is clustering, that is, dividing the overall area into subareas so that all the individuals in each subarea are concentrated in a particular area. This has a direct relation to graph theory in mathematics. Roughly speaking, a graph consists of some points and lines connecting them. We say that a graph is connected if, for any two points, we can find a way to get from one to the other via the lines of the graph. Therefore, any graph can be viewed as a union of connected graphs called its connected components. If we consider individuals as points and a line connects two points if their distance is less than a certain value, then we have the graph of individuals for a certain distance value. Now clusters are just the connected component of the graph. Note that this kind of clustering is an equivalent way of looking at the well-known distribution clustering since the individuals that are close to each other are more likely to have the same distribution. This kind of clustering depends entirely on the initial distance we consider. But it does give us a way to compare two species across distances. For BCI, since the area coverage of an individual is on average about 2 square meters, we use uniform multipliers of $\sqrt{5}$ to identify lines in the graph. Figure \ref{C1} shows the clustering of the species rinosy considering different even multipliers of $\sqrt{5}$. Different colors correspond to cluster classes of different individuals. We choose this species because cluster classification can be easily visualized by distances.

\begin{figure}[!h]
\centering
\includegraphics[width=0.48\linewidth]{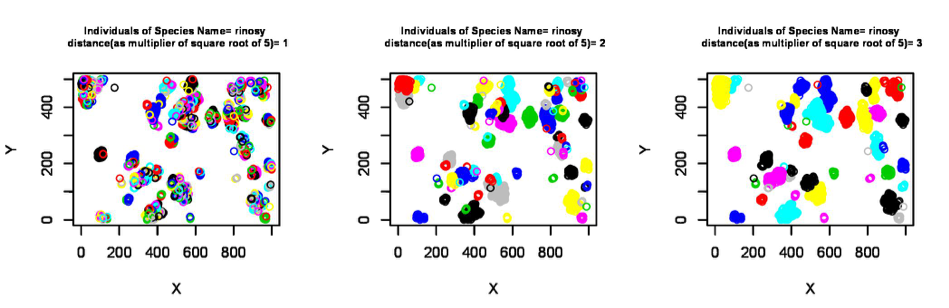}
\includegraphics[width=0.48\linewidth]{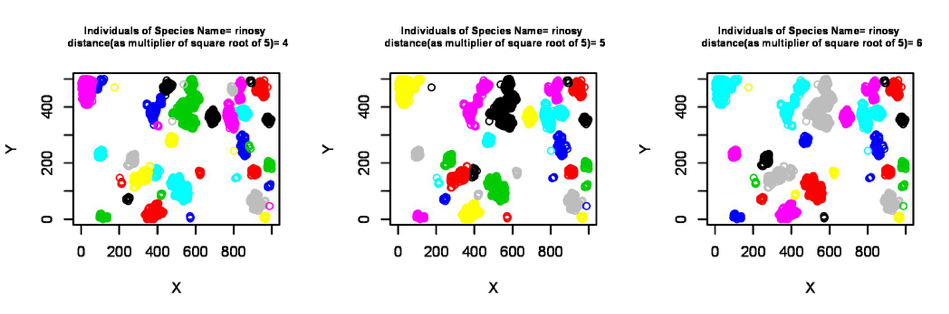}
\includegraphics[width=0.48\linewidth]{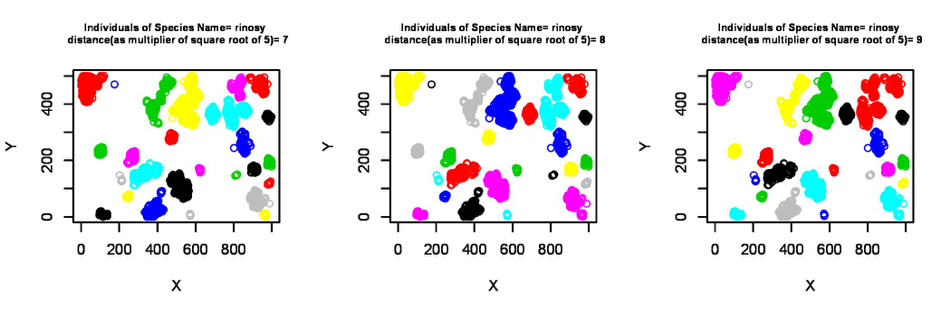}
\includegraphics[width=0.48\linewidth]{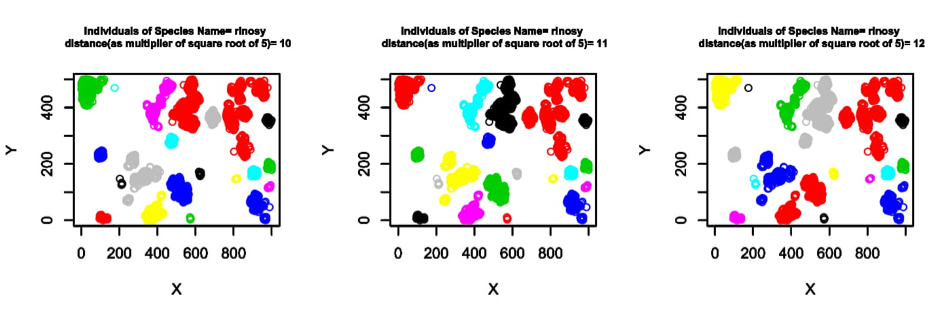}
\includegraphics[width=0.48\linewidth]{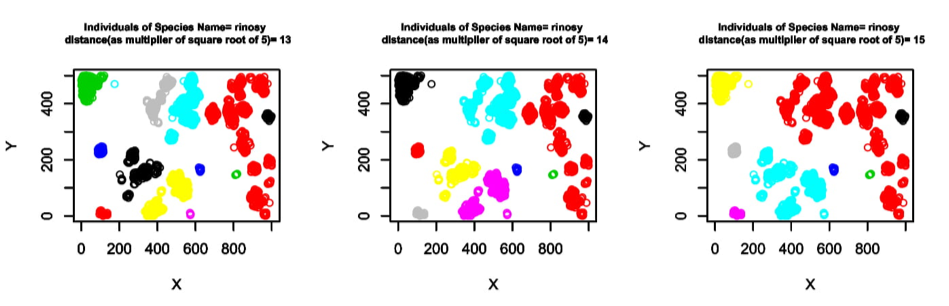}
\includegraphics[width=0.48\linewidth]{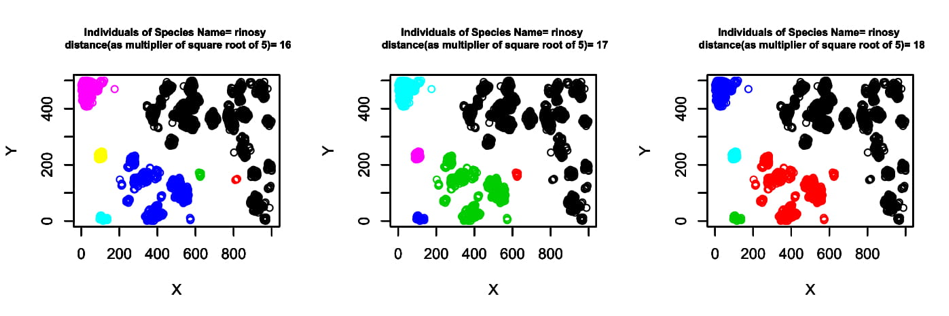}
\includegraphics[width=0.48\linewidth]{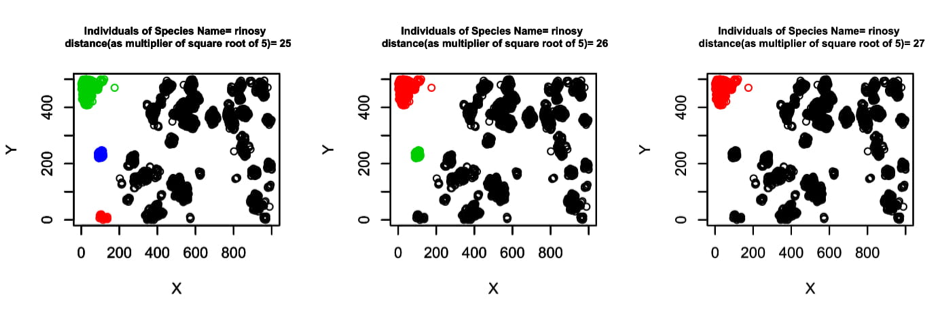}
\includegraphics[width=0.48\linewidth]{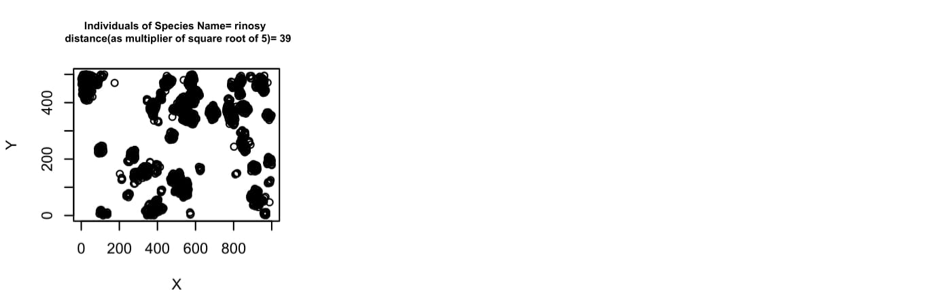}
\caption{In the BCI data and for the species rinosy, we show how individuals are clustered as a function of distances between individuals.}
\label{C1}
\end{figure}

\begin{figure}[!h]
\includegraphics[width=0.48\linewidth]{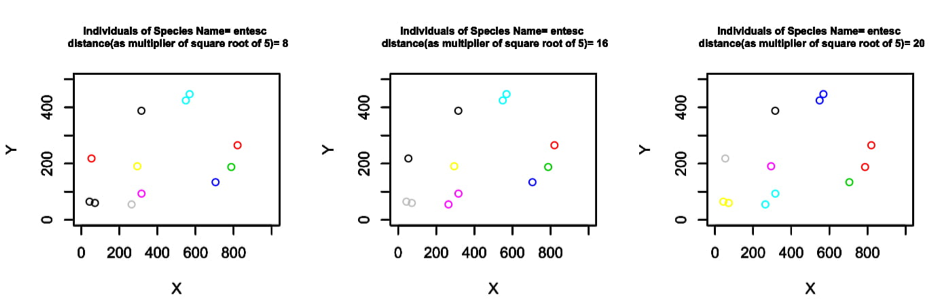}
\includegraphics[width=0.48\linewidth]{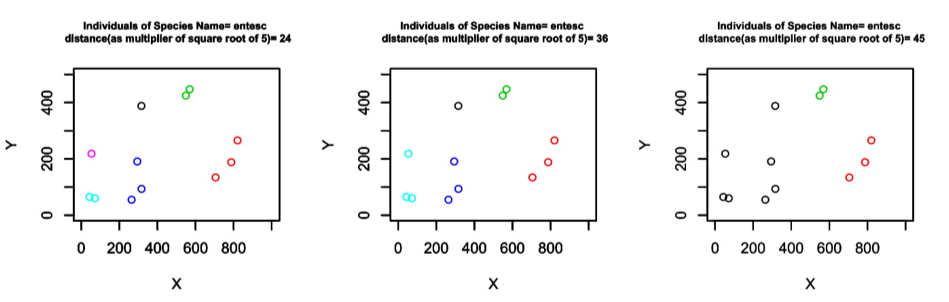}
\includegraphics[width=0.48\linewidth]{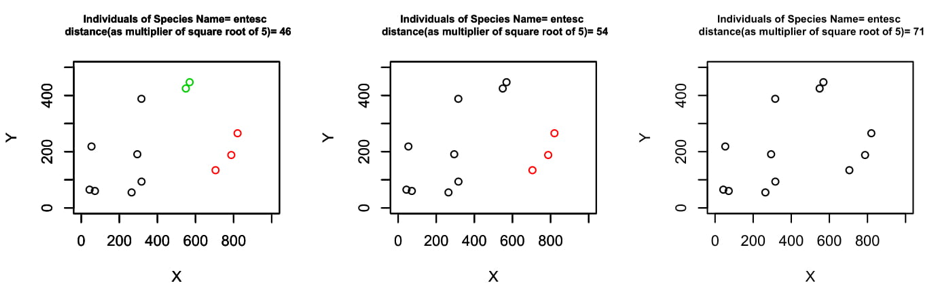}
\caption{In the BCI data and for the species entesc, it shows how individuals are clustered as a function of the distances between individuals.}
\label{C2}
\end{figure}

\begin{figure}[!h]
\centering
\includegraphics[width=0.48\linewidth]{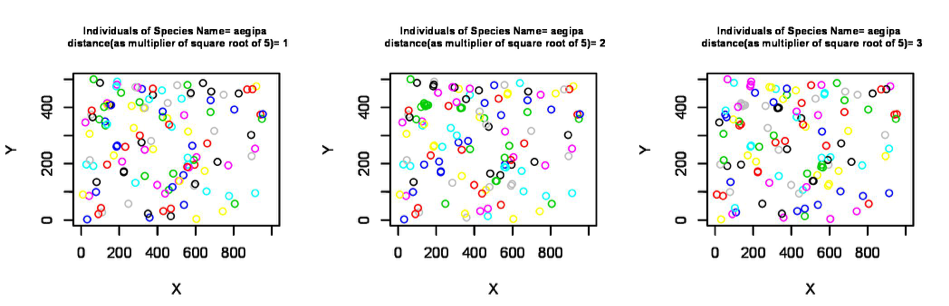}
\includegraphics[width=0.48\linewidth]{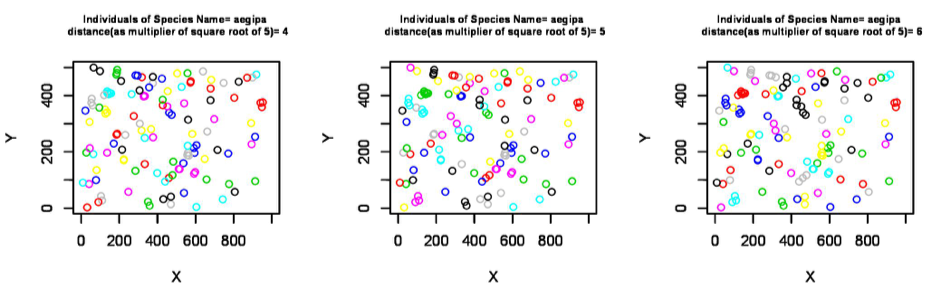}
\includegraphics[width=0.48\linewidth]{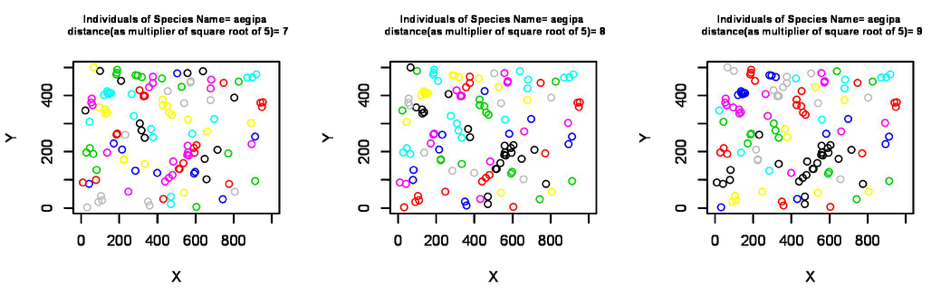}
\includegraphics[width=0.48\linewidth]{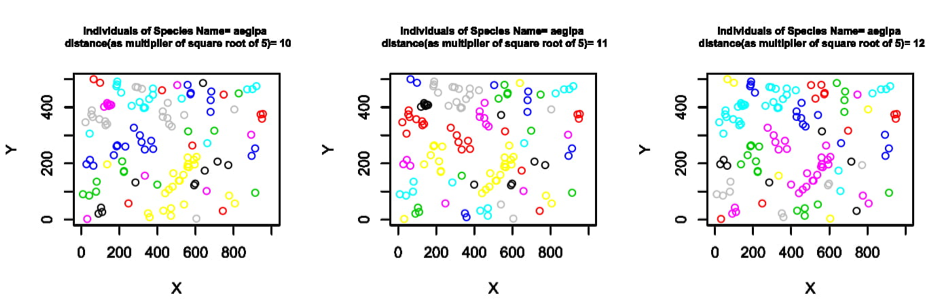}
\includegraphics[width=0.48\linewidth]{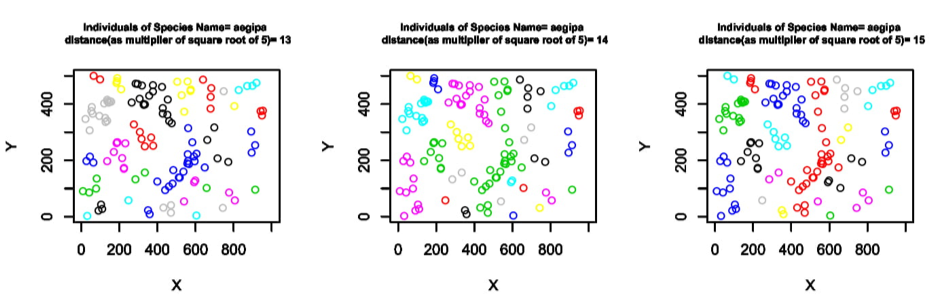}
\includegraphics[width=0.48\linewidth]{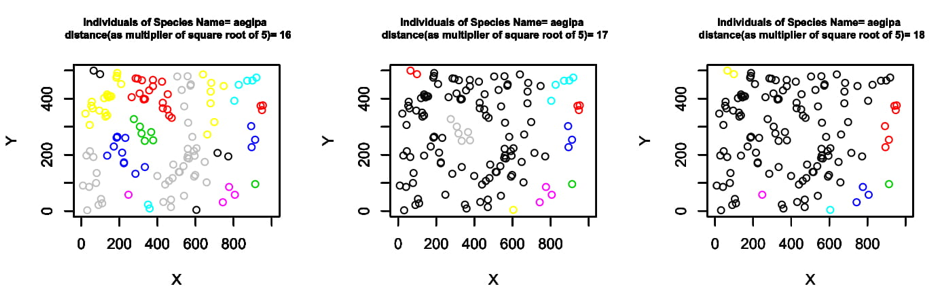}
\includegraphics[width=0.48\linewidth]{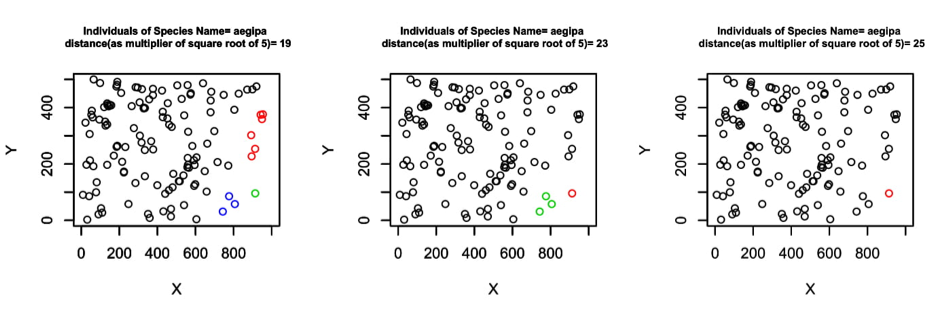}
\includegraphics[width=0.48\linewidth]{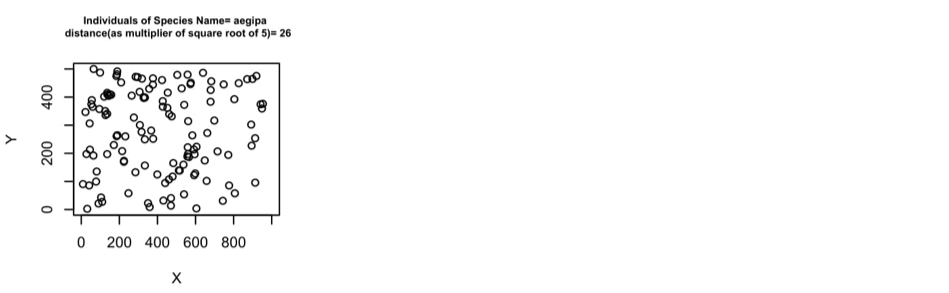}
\caption{In the BCI data and for the species aegipa, it is shown how individuals are clustered as a function of the distances between individuals.}
\label{C3}
\end{figure}

\cite{Plotkin2002} used the idea of critical distance, which is the distance at which smaller than this the number of classes of clusters of individuals will be high, but larger than this it will be small. However, the concept of critical distance cannot be easily calculated because the values for high or low are not fixed. See Figure \ref{C2} for the clustering of species entesc in BCI, considering different uniform multipliers of $\sqrt{5}$. As we can see, the critical distance is not reachable for the individuals of this species (because it is widely distributed), and by slightly increasing the distance, we slightly reduce the number of classes that cause the critical distance to be unreachable. Similarly, for clustering the species aegipa, which has 126 individuals, we consider a different uniform multiplier of $\sqrt{5}$, see Figure \ref{C3}.


\begin{figure}[!h]
\centering
\includegraphics[width=0.6\linewidth]{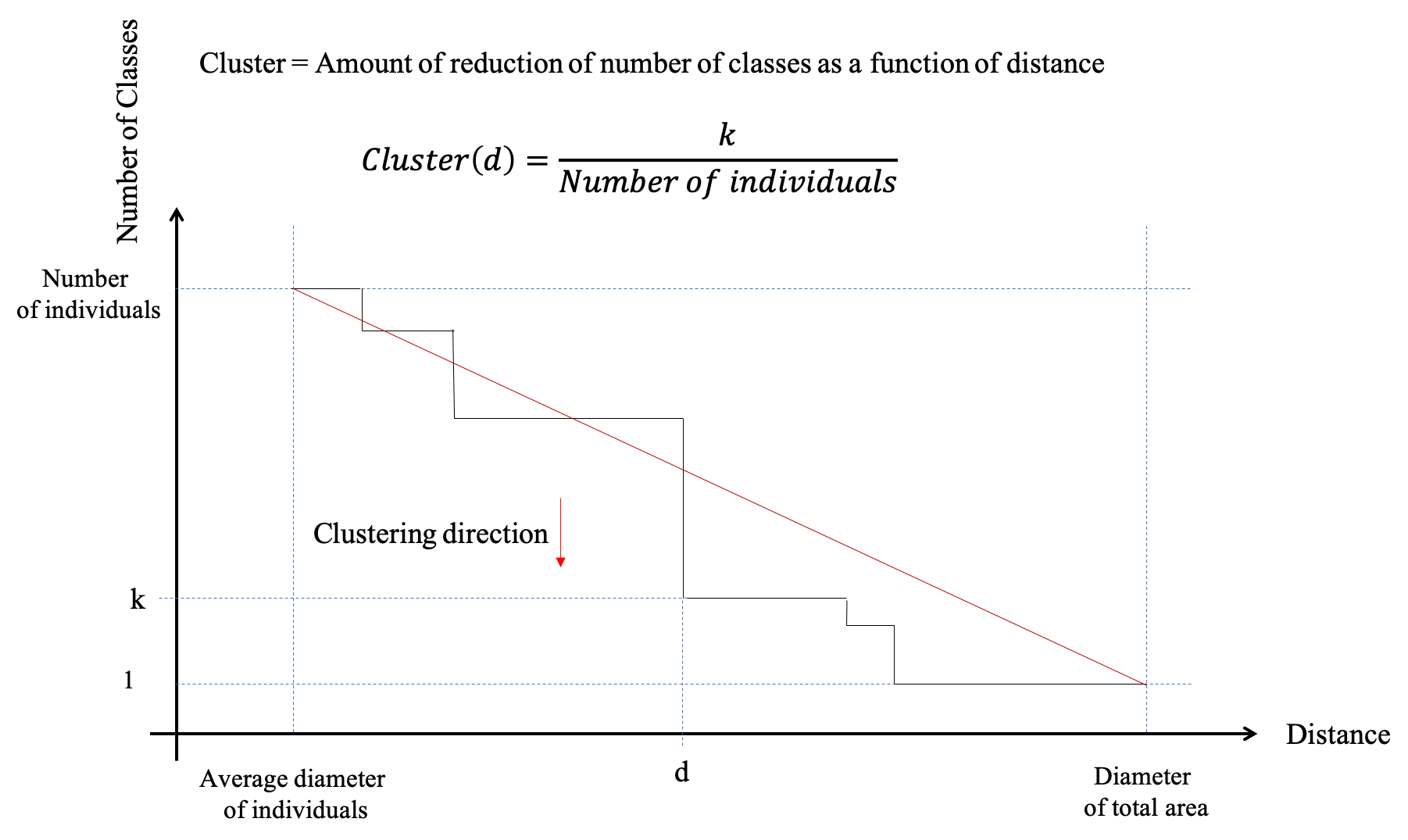}
\caption{The meaning of cluster and cluster as a function of distance, allowing us to compare different species.}
\label{C4}
\end{figure}

We propose instead to classify clustering by using different distances instead of a single numerical value. We know that for each species, there is a distance at which all individuals are in the same class that is at most the diameter of the area, and there is also a distance at which each class consists of exactly one individual (the average diameter of the individuals). If we increase the distance from the smallest to the largest distance, then the number of classes reduces from the total number of individuals to 1 for each of the smallest and largest distances. The faster the reduction as a function of distance, the stronger the clustering at that distance. This, therefore, allows species to be compared, and for a fixed distance, if the proportion of the number of class to the total number of individuals is lower, it means that the species is more clustered with respect to distance, see Figure \ref{C4}, for the cluster as a function of distance.

\end{document}